%% file: main.tex
\documentclass[sigconf]{acmart}
\AtBeginDocument{%
  }

\usepackage{pdflscape}
\usepackage{subcaption}
\usepackage{multirow}
\usepackage{balance}
\usepackage{float}
\usepackage{cuted}
\copyrightyear{2026}
\acmYear{2026}
\setcopyright{cc}
\setcctype{by}
\acmConference[KDD '26]{Proceedings of the 32nd ACM SIGKDD Conference on Knowledge Discovery and Data Mining V.2}{August 09--13, 2026}{Jeju Island, Republic of Korea}
\acmBooktitle{Proceedings of the 32nd ACM SIGKDD Conference on Knowledge Discovery and Data Mining V.2 (KDD '26), August 09--13, 2026, Jeju Island, Republic of Korea}
\acmDOI{10.1145/3770855.3818898}
\acmISBN{979-8-4007-2259-2/2026/08}
\begin{document}

\title[GraphCliff: Short--Long Range Gating for Activity Cliffs]{GraphCliff: Short–Long Range Gating for Modeling Critical Activity Changes Caused by Subtle Molecular Differences}

%% ACM requires a valid ORCID for every author (verified against orcid.org).
\author{Hajung Kim}
\orcid{0000-0003-3209-3494}
\affiliation{%
  \institution{Korea University}
  \city{Seoul}
  \country{South Korea}
}
\email{hajungk@korea.ac.kr}

\author{Jueon Park}
\orcid{0000-0002-0613-120X}
\affiliation{%
  \institution{Korea University}
  \city{Seoul}
  \country{South Korea}}
\email{jueon_park@korea.ac.kr}

\author{Junseok Choe}
\orcid{0000-0001-9548-7146}
\affiliation{%
  \institution{Korea University}
  \city{Seoul}
  \country{South Korea}}
\email{juns94@korea.ac.kr}

\author{Seungheun Baek}
\orcid{0000-0002-3231-478X}
\affiliation{%
  \institution{Korea University}
  \city{Seoul}
  \country{South Korea}}
\email{sheunbaek@korea.ac.kr}

\author{Hyeon Hwang}
\orcid{0009-0003-7851-3834}
\affiliation{%
  \institution{Korea University}
  \city{Seoul}
  \country{South Korea}}
\email{hyeon-hwang@korea.ac.kr}

%% Per ACM policy, multiple affiliations must each be in their own
%% \affiliation{} block (not batched in one \institution string), and each
%% must carry its own city/country.
\author{Jaewoo Kang}
\authornote{Corresponding author.}
\orcid{0000-0001-6798-9106}
\affiliation{%
  \institution{Korea University}
  \city{Seoul}
  \country{South Korea}}
% \affiliation{%
%   \institution{AIGEN Sciences}
%   \city{Seoul}
%   \country{South Korea}}
\email{kangj@korea.ac.kr}

\renewcommand{\shortauthors}{Hajung Kim et al.}

\begin{abstract}
 The quantitative structure–activity relationship assumes a smooth mapping between molecular structure and biological activity. However, activity cliffs, defined as pairs of structurally similar compounds with large potency differences, break this continuity. Recent activity cliff benchmarks show that machine learning models with extended connectivity fingerprints outperform graph neural networks. Our analysis shows that the embedding distances of conventional graph neural networks fail to reflect the activity differences, collapsing structurally similar yet functionally divergent molecules into nearly indistinguishable representations. To recover sensitivity to such local changes while preserving global molecular context, we propose GraphCliff, which integrates short- and long-range information at the node level through a locally conditioned gating mechanism. Experimental results demonstrate that GraphCliff consistently improves performance on both non-cliff and cliff compounds, with layer-wise embedding analyses attributing these gains to sharper discrimination of structurally similar molecules relative to strong baseline graph models.
\end{abstract}

%% IMPORTANT: regenerate this block at https://dl.acm.org/ccs and paste the
%% exact XML + \ccsdesc lines here (the numeric concept_id values below are a
%% best-effort starting point and must be confirmed with the ACM CCS tool,
%% which you also need for the mandatory CCS field on the submission page).



% \received{1 February 2026}
% \received[revised]{1 February 2026}

\maketitle

\input{0_introduction}

\input{1_related_works}
\input{2_methods}
\input{3_results}
\input{4_analysis}

\input{5_conclusion}

\section*{Dataset and Code Availability}
The code and datasets are available at \href{https://github.com/dmis-lab/GraphCliff}{https://github.com/dmis-lab/GraphCliff}.

\section*{Ethical Considerations}
This study does not involve human subjects and does not require IRB approval. All molecular datasets used are publicly available. The method is intended for research and decision support purposes only. During the preparation of this manuscript, the authors used generative AI
to assist with language editing and clarity improvements.
All technical content and results were verified by the authors.

\section*{Funding}
This work was supported by the National Research Foundation of Korea (NRF) grant funded by the Korea government (MSIT and MOE) (No. RS-2025-16652968), NRF-2023R1A2C3004176, RS-2023-00262002, and HR20C002103. This work was also supported by the ICT Creative Consilience Program through the Institute of Information \& Communications Technology Planning \& Evaluation (IITP) grant funded by the Korea government (MSIT) (IITP-2026-RS-2020-II201819). MSIT denotes the Ministry of Science and ICT, and MOE denotes the Ministry of Education.

\clearpage
\bibliographystyle{ACM-Reference-Format}
\balance
\bibliography{ref}

\clearpage

%% Appendix layout:
%%  - Page 1 (one-column): section heading + the wide per-target results table.
%%    One-column lets the heading and the full-width table share a page; the
%%    table is pinned in place with [H] (float package) so it cannot drift.
%%  - Page 2 (two-column): the two full-width figures (figure*) on top and the
%%    error-analysis prose flowing beneath them in two columns.
\onecolumn
\appendix
\section{Appendix}

\input{tables/full_results}

\input{6_appendix}

\end{document}

%% file: 0_introduction.tex
\section{Introduction}

Quantitative Structure–Activity Relationship (QSAR) is based on the premise that molecules with similar structures have similar biological activity. QSAR modeling plays a crucial role in drug discovery by reducing the number of compounds that require experimental testing, supporting tasks such as virtual screening for hit identification, lead optimization, and ADMET evaluation~\citep{cherkasov2014qsar}. To this end, a wide range of machine learning and deep learning models have been developed to predict molecular properties and biological activities directly from molecular structures~\citep{hu2019strategies, wang2022molecular, heid2023chemprop, li2023knowledge, qiao2025self}.

However, there exists a class of cases that breaks the continuity of the typical structure–activity relationship, known as \textit{activity cliffs}. Unlike the conventional assumption that structurally similar molecules exhibit similar activities, activity cliffs describe cases where minor structural differences lead to large and abrupt changes in activity. They are formally quantified as the ratio of the activity difference between two compounds to their distance in a given chemical space~\citep{maggiora2006outliers}. In practical terms, activity cliffs are defined as pairs or groups of structurally similar compounds that are active against the same target protein but exhibit large potency differences~\citep{stumpfe2019evolving}. Although analog groups corresponding to activity cliffs may deviate from general QSAR assumptions, they highlight the importance of local structural changes and provide valuable insight into processes such as hit-to-lead optimization and structural alert development~\citep{stumpfe2012exploring, wedlake2019structural}.

\begin{figure*}[htbp]
    \centering
    \includegraphics[width=\textwidth]{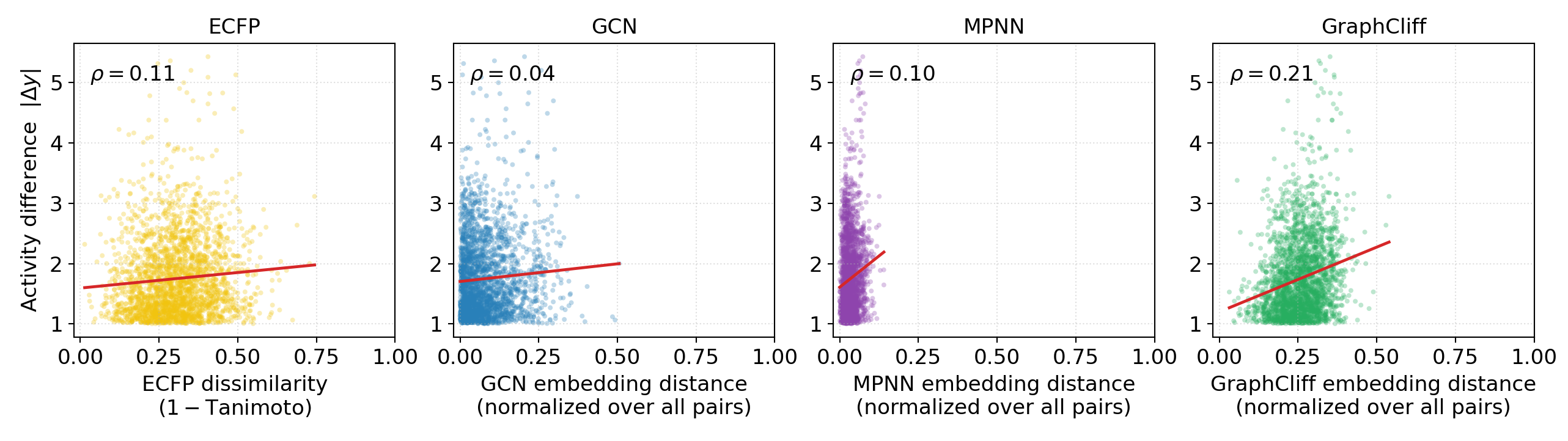}
    \caption{For each activity cliff pair in the CHEMBL244 (Ki) dataset, the activity difference $|\Delta y|$ ($y= - \log_{10}\!\left(K_i\,[\mathrm{nM}]\right)$) versus the pairwise distance in each representation: ECFP dissimilarity ($1-$Tanimoto) and graph-embedding Euclidean distance (min--max normalized over all molecule pairs). $\rho$ denotes the Spearman rank correlation. GraphCliff's embedding distance tracks activity differences most strongly, whereas fixed fingerprints (ECFP) and conventional GNNs (GCN, MPNN) show weak correlation.}
    \label{fig:embedding_comparison}
\end{figure*}

Motivated by the importance of activity cliffs in drug discovery, the MoleculeACE~\citep{van2022exposing} dataset is curated from ChEMBL~\citep{gaulton2012chembl} and used to evaluate a wide range of models. The results revealed that machine learning models with extended connectivity fingerprints (ECFPs) consistently outperform deep learning approaches, with CNN and LSTM using SMILES~\citep{weininger1988smiles} providing moderate success, while Transformer and GNNs generally underperformed. The strong performance of ECFPs stems from their design, in which binary bit vectors encode radius-based substructures that are highly sensitive to chemical modifications~\citep{rogers2010extended}. Hashing atom-centered neighborhoods into sparse vectors yields a strong inductive bias and low variance, allowing ECFP-based models to generalize more reliably than flexible deep models in small-data regimes.

In contrast, GNNs introduce numerous parameters and high modeling flexibility, which increase variance under limited data~\citep{baptista2022evaluating}. Moreover, as layers deepen, node embeddings become homogenized due to Laplacian smoothing, leading to the over-smoothing phenomenon where fine-grained local distinctions vanish~\citep{wu2023demystifying}. This explains why LSTM and CNN models, which emphasize local structural changes, often perform better than Transformer and GNNs. Nevertheless, molecular graphs inherently preserve rich structural information, where atoms are represented as nodes and bonds as edges, with extensions to 3D coordinates, charges, or bond orders directly incorporated~\citep{kearnes2016molecular}. Unlike ECFPs that rely on predefined radius-based hashing, graph representations can adaptively capture complex topological patterns and long-range dependencies. The central challenge is therefore to design graph architectures that combine the expressiveness and long-range modeling capacity of molecular graphs with sensitivity to local patterns, without suffering from over-smoothing.

To confirm this difficulty, we examined how faithfully different molecular representations encode activity cliffs on CHEMBL244 (Ki), a representative cliff-rich target. For each activity cliff pair, we compared the activity difference \(|\Delta y|\) with the pairwise distance in representation space, using ECFP dissimilarity (\(1-\mathrm{TanimotoSimilarity}\)) for fingerprints and Euclidean distance~\citep{liberti2014euclidean} for graph embeddings, normalized over all molecule pairs within CHEMBL244 (Ki). An effective representation should place cliff pairs with larger activity differences proportionally farther apart. Consistent with the local sensitivity of fingerprints, ECFP dissimilarity is widely spread across cliff pairs (Figure~\ref{fig:embedding_comparison}). However, it correlates only weakly with the activity difference (Spearman \(\rho=0.11\)), so structural dissimilarity alone does not capture the \emph{magnitude} of the activity change. Conventional GNNs are worse. They compress cliff pairs into a narrow band of small embedding distances (GCN \(\rho=0.04\), MPNN \(\rho=0.10\)), failing to separate them at all. Thus, even when fingerprints perform well, the embedding geometry of existing representations does not directly reflect how much activity changes across a cliff.

This limitation motivates graph models whose embedding geometry better reflects the activity differences that define cliffs. We introduce \textbf{GraphCliff}, which uses a node-level gating mechanism to selectively integrate short-range local features with long-range molecular context. As shown in Figure~\ref{fig:embedding_comparison}, GraphCliff avoids this collapse and yields embeddings whose distances track activity differences most strongly among all representations (\(\rho=0.21\), roughly twice the strongest baseline), indicating sharper sensitivity to the local structural changes that drive activity cliffs.

While integrating local and global information is itself well studied in graph learning (multi-hop, multi-scale, and spectral GNNs), we do not claim novelty in the individual operators we employ. Our contribution is a \textbf{problem-driven specialization} for activity cliffs. The setting demands sensitivity to a minimal local perturbation (one atom or bond) while still encoding the broader context that determines its functional consequence. These two signals conflict, because aggressive global propagation washes out the local distinctions that define cliffs. GraphCliff resolves this tension through a \textbf{node-level, locally conditioned gate}. A single GINE projection is split into three streams, one conditioning the gate on each atom's local neighborhood, one carrying the long-range spectral signal, and one providing an unconditional local bypass. Unlike prior multi-scale methods that mix local and global signals uniformly across nodes, this lets each atom adaptively decide how much global context to admit. Our contributions are as follows:
\begin{itemize}
    \item We identify a tension specific to activity cliff prediction, the simultaneous need for \textbf{local perturbation sensitivity and global contextualization}, and argue that it requires an architecture that explicitly separates and adaptively balances these two information streams rather than uniformly integrating them.
    \item We introduce \textbf{GraphCliff}, a graph architecture that decomposes a short-range embedding into three streams and fuses short- and long-range representations through a \textbf{node-level, locally conditioned gate}, directly targeting the loss of local sensitivity and over-smoothing observed in existing GNNs.
    \item We provide extensive empirical evidence on the benchmark, demonstrating \textbf{consistent improvements on both non-cliff and activity cliff compounds}, and a comprehensive analysis showing that our model mitigates over-smoothing and yields \textbf{more discriminative representations} than existing GNNs.
\end{itemize}

%% file: 1_related_works.tex
\section{Related Work}
\paragraph{Contextual Dependencies Across Ranges.}
Modeling dependencies at different contextual ranges is important when both fine local detail and broad structural context are needed. In sequence modeling, StripedHyena2~\citep{ku2025systems} addresses this by decomposing information into short-, medium-, and long-range components with specialized modules, then integrating them through learnable gating. Although designed for 1D token sequences, this principle is closely aligned with molecular graph prediction, where local neighborhoods often determine activity cliffs while longer-range structure provides broader QSAR context. GraphCliff adapts this idea by combining short-range message passing with long-range propagation through a node-level gate.

\paragraph{Multi-scale Propagation and Gating in GNNs.}
Combining information across propagation ranges is a central theme in graph learning. Multi-hop and decoupled-propagation models such as APPNP and GCNII~\citep{gasteiger2018predict, chen2020simple} propagate information over many hops while reducing over-smoothing, multi-scale aggregation methods such as JK-Net~\citep{xu2018representation} combine embeddings from different layers, and spectral filters such as ChebNet~\citep{hammond2011wavelets} capture multi-hop context through polynomial Laplacian filters within a single layer. Gating mechanisms have also been used to control information flow: GGS-NN~\citep{li2015gated} stabilizes recurrent node updates, GaAN~\citep{zhang2018gaan} weights attention heads, and GFGN~\citep{jin2021graph} modulates feature channels. These approaches broaden receptive fields or improve information control, but they do not explicitly distinguish short-range and long-range structural signals at the node level. GraphCliff instead uses each atom's local representation to gate how much long-range signal is admitted, while preserving local information through a bypass connection.

\paragraph{Activity Cliffs.}
Activity cliffs refer to pairs of compounds that are structurally similar but exhibit large potency differences~\citep{stumpfe2019evolving}. Early studies formalized this phenomenon using Matched Molecular Pairs, where small substitutions lead to large changes in activity. MoleculeACE~\citep{van2022exposing} provides a dedicated ChEMBL-derived benchmark for activity cliff prediction, defining cliffs using complementary substructure, scaffold, and SMILES similarity criteria together with at least a 10-fold difference in $K_i$. The benchmark evaluates diverse model families, including graph models such as AFP~\citep{xiong2019pushing}, GAT~\citep{velivckovic2017graph}, GCN~\citep{kipf2016semi}, and MPNN~\citep{gilmer2017neural}; sequence and descriptor models such as CNN~\citep{kimber2021maxsmi}, LSTM~\citep{hochreiter1997long}, MLP, and Transformer~\citep{vaswani2017attention}; and classical machine-learning methods such as GBM~\citep{friedman2001greedy}, KNN~\citep{fix1985discriminatory}, RF~\citep{predictors1996bagging}, and SVM~\citep{cristianini2000shawe}. Notably, this benchmark revealed that classical fingerprint-based methods often outperform graph models, exposing the difficulty current GNNs face with the localized structural changes that define activity cliffs.

%% file: 2_methods.tex
\begin{figure*}[htbp]
    \centering
    \includegraphics[width=1\textwidth]{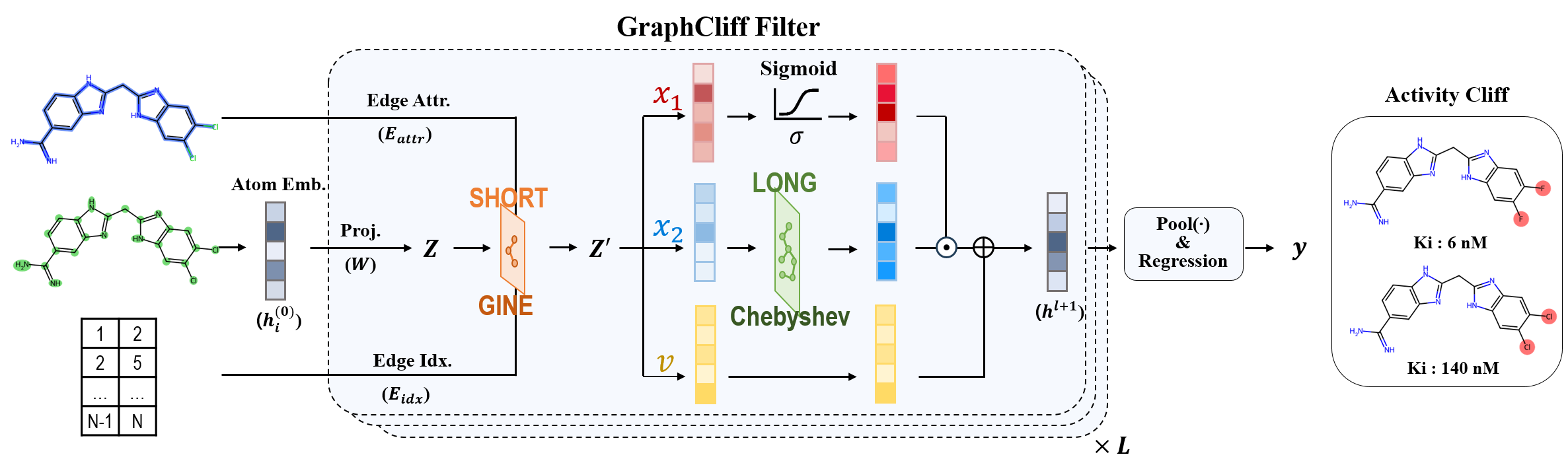}
    \caption{Overall architecture of GraphCliff.}
    \label{fig:main_architecture}
\end{figure*}

\section{Methods}
\subsection{Datasets}

We employed MoleculeACE as the benchmark dataset, which consists of curated compound–protein interaction data extracted from ChEMBL. Each dataset contains experimentally measured potency values (\(K_i\)) for compounds targeting a specific protein, with low-quality samples removed during curation according to predefined criteria. In total, MoleculeACE comprises 30 datasets, each corresponding to a distinct protein target, and \(K_i\) values are used as regression labels to evaluate model performance across diverse biological contexts.

Activity cliff pairs in these datasets were defined based on two criteria: structural similarity and a large potency discrepancy. Structural similarity between compounds was assessed using three complementary measures: substructure similarity, scaffold similarity, and SMILES string similarity. Two compounds were considered structurally similar if at least one of these similarity scores exceeded 0.9. If their potency values differed by at least a tenfold change, the pair was regarded as exhibiting a significant activity difference. Compound pairs satisfying both criteria were designated as activity cliffs. In the benchmark, activity cliff information is encoded in a binary manner, indicating whether a compound belongs to any cliff, without explicitly enumerating specific cliff pairs. We use the official MoleculeACE train/test split exactly as provided. Each dataset CSV contains a \texttt{split} column with pre-assigned \texttt{train}/\texttt{test} labels, which we load directly without any re-splitting. This split is stratified to preserve the proportion of activity cliff compounds and to keep structurally similar molecules from being divided across train and test, preventing leakage by construction. The binary \texttt{cliff\_mol} indicator is used solely for evaluation (to compute $\mathrm{RMSE}_{\text{cliff}}$). It is never provided as a training signal, so GraphCliff has no access to cliff labels during training.

\subsection{Model}

We take inspiration from StripedHyena2, which decomposes its core operator into short-, middle-, and long-range variants that are sequentially connected to capture information across multiple scales. Extending this design principle to molecular graphs, we construct a graph architecture where conventional graph modules are adapted to process short- and long-range dependencies and their outputs are fused through a learnable, node-level sigmoid gate. An overview of this architecture is illustrated in Figure~\ref{fig:main_architecture}, and the formulation of our model is presented in the following equations.

\paragraph{\textbf{Problem setup and notation}}
We represent each molecule as a graph $G = (V, E)$, where $V$ is the set of atoms and $E$ is the set of chemical bonds. Let $N = |V|$ and $M = |E|$ denote the numbers of atoms and bonds, respectively. Each atom is associated with an input feature vector, forming the node feature matrix $x \in \mathbb{R}^{N \times d_{\text{in}}}$. Bond connectivity is represented by the edge index $E_{idx} \in \mathbb{R}^{M \times 2}$, and each bond carries an edge attribute vector stored in $E_{attr} \in \mathbb{R}^{M \times d_{attr}}$. Given a molecular graph $G$, the learning objective is to predict a target property $y$ through a parametric mapping $f(G;\theta)$. Unless otherwise stated, we use $d$ to denote the hidden dimensionality of intermediate node embeddings.

\paragraph{\textbf{Atom encoding}}
Given the node feature matrix $x$, the model first maps each atom to a $d$-dimensional hidden representation via $h^{(0)} = \phi_{\text{atom}}(x) \in \mathbb{R}^{N \times d}$, where $\phi_{\text{atom}}$ denotes an MLP followed by normalization and a nonlinear activation function.

\paragraph{\textbf{GraphCliff filter}}
Inspired by StripedHyena2, in which a single projection is decomposed into multiple functional streams, each GraphCliff filter maps \(d\)-dimensional node representations into a \(3d\)-dimensional space and integrates short- and long-range information through a gating mechanism. The expanded representation is first processed by GINE~\citep{xu2018powerful} as a short-range path that captures local neighborhood interactions, and its output is split along the feature dimension into three node-level streams used for long-range propagation, gate conditioning, and local bypass. The long-range stream is then processed by a Chebyshev polynomial filter~\citep{hammond2011wavelets}, which captures multi-hop dependencies within a single layer and thereby avoids the need to stack multiple GNN layers. Recent work demonstrates that Chebyshev polynomials operate directly on the normalized Laplacian, propagating information across multiple hops without explicit edge rewiring or architectural modifications that distort the original topology~\citep{hariri2025chebnet}. 

At layer \(\ell\), given hidden node representations \( h^{(\ell)} \in \mathbb{R}^{N \times d} \), we first apply normalization followed by a linear projection:
\begin{equation}
Z = h^{(\ell)}W, \quad Z \in \mathbb{R}^{N \times 3d},
\end{equation}
where \( W \in \mathbb{R}^{d \times 3d} \) is a trainable projection matrix. The short-range path applies a GINE message passing operator to the projected features \( Z \):
\begin{equation}
Z' = \text{GINE}(Z, E_{idx}, E_{attr}),
\end{equation} where $E_{idx}$ denotes edge index and $E_{attr}$ denotes edge features. The GINE operator is defined as:
\begin{equation}
z'_i = \psi\left((1+\epsilon) z_i + 
\sum_{j \in \mathcal{N}(i)} \left(z_j + \phi(e_{ij})\right)\right),
\end{equation}

where $e_{ij}$ denotes the edge attribute associated with the directed edge from source node $j$ to target node $i$, $\phi$ is an MLP applied to edge attributes, $\psi$ is a node-wise MLP, and $\epsilon$ is a learnable scalar parameter. Since $Z \in \mathbb{R}^{N \times 3d}$, each node embedding $z_i$ and $z_j$ lies in $\mathbb{R}^{3d}$, and the edge MLP $\phi$ maps $e_{ij}$ into the same $3d$-dimensional space to ensure dimensional consistency in the aggregation term. The GINE output \( Z' \in \mathbb{R}^{N \times 3d} \) is then split along the feature dimension into three node-level streams:
\begin{equation}
Z' = [\,x_2 \;\|\; x_1 \;\|\; v\,], \quad x_2, x_1, v \in \mathbb{R}^{N \times d}.
\end{equation}
Here, \(x_2\) is passed to the long-range path, \(x_1\) conditions the node-level gate, and \(v\) serves as a local bypass that preserves short-range information.

To capture global context, we define the long-range function \(\mathrm{Long}(\cdot)\) as a Chebyshev polynomial filter applied to \(x_2\). Specifically, we compute Chebyshev polynomials over the normalized adjacency matrix \( \hat{A} \):
\begin{equation}
T_0 = x_2, \quad T_1 = \hat{A}x_2, \quad T_k = 2\hat{A}T_{k-1} - T_{k-2} \ (k \ge 2).
\end{equation}
The long-range function then computes a learnable weighted combination of these polynomial bases:
$\mathrm{Long}(x_2)=\sum_{k=0}^K \alpha_k T_k$, 
where $\alpha_k$ are learnable coefficients. Thus, \(\mathrm{Long}(x_2)\) aggregates information from multiple hop ranges in a single filter layer without repeatedly applying message passing.

We combine short- and long-range information using a sigmoid gating function:
\begin{equation}
g = \sigma(x_1), \qquad 
u = g \odot \mathrm{Long}(x_2) + v,
\end{equation}
where \( \sigma \) denotes the element-wise sigmoid function and \( \odot \) is the element-wise product. Gating mechanisms have been shown to alleviate over-smoothing in GNNs by adaptively regulating information flow~\citep{xin2020graph}, providing empirical support for our design. Finally, the filter layer output is updated via a residual connection as 
$h^{(\ell+1)} = u^{(\ell)} + h^{(\ell)}$. We stack \(L\) GraphCliff filters sequentially, where the output of each layer serves as the input to the next:
\begin{equation}
h^{(\ell+1)} = \mathrm{GraphCliffFilter}^{(\ell)}\!\big(h^{(\ell)}, E_{idx}, E_{attr}\big),
\qquad \ell = 0, \dots, L-1.
\end{equation}

\paragraph{\textbf{Pooling and regression}} 
Finally, we apply an attention-based graph pooling operation to adaptively select and aggregate informative nodes. Specifically, we employ SAGPool~\citep{lee2019self}, and the resulting pooled representation is passed to a regression head to produce the final output corresponding to the target property.
Formally, given the final-layer node embeddings \(h^{(L)}\), SAGPool produces a graph-level representation that the regression MLP \(\phi_{\text{reg}}\) maps to the predicted property:
\begin{equation}
\hat{y} = \phi_{\text{reg}}\!\left(\text{SAGPool}\!\left(h^{(L)}\right)\right).
\end{equation}

\input{tables/main_results}

\paragraph{\textbf{Implementation details}}
Unless otherwise stated, all hyperparameters are fixed across the 30 datasets to ensure a consistent evaluation. We stack $L=3$ GraphCliff filters with hidden dimension $d=256$ and use a $3d$ projection for the three functional streams. The Chebyshev polynomial order is fixed at $K=3$ as a single default, without per-dataset tuning. This captures up to 3-hop context within one layer and covers a substantial portion of MoleculeACE molecules, whose median graph diameter is roughly ten bonds. SAGPool is configured with a pooling ratio of $0.8$ and uses SAGEConv as the scoring network. Its attention scores are learned independently of the gating mechanism and serve only the graph-level readout, whereas the gate controls local/global integration within each GraphCliff filter.

\subsection{Theoretical Analysis}
\label{sec:theory}
We formalize why standard message passing struggles on activity cliffs and how GraphCliff alleviates it by separating local and long-range streams and recombining them through a node-level gate.

\paragraph{\textbf{Why standard message passing fails.}}
Activity cliffs hinge on a minimal structural change producing a large activity change, yet standard message passing exponentially suppresses this signal. For linearized propagation $h^{(L)} = \hat{A}^{L} X W$, the normalized adjacency $\hat{A}$ has eigenvalues $1 = \lambda_1 \ge |\lambda_2| \ge \dots$. For a connected graph $|\lambda_i| < 1$ ($i \ge 2$), so $\hat{A}^{L}$ converges to rank one and node embeddings collapse toward the leading eigenvector. For a cliff pair $M_1, M_2$ differing at a single atom $v$, the mean-pooled difference $\Delta h_G^{(L)} = \frac{1}{N}\sum_i (h_i^{(L)}(M_1) - h_i^{(L)}(M_2))$ decays as $\mathcal{O}(|\lambda_2|^{L})$, i.e., exponentially in depth. This is precisely the over-smoothing that we quantify via decreasing MAD and Dirichlet energy in Section~\ref{sec:oversmoothing_analysis}.

\paragraph{\textbf{How gating alleviates smoothing.}}

% GraphCliff avoids this collapse by routing local information through a path that bypasses long-range propagation.

GraphCliff alleviates this collapse through two mechanisms: a local bypass that preserves short-range perturbation signals, and a node-level gate that controls how much long-range information is admitted at each atom. Each layer outputs $u_v = \sigma(x_{1,v}) \odot \mathrm{Long}(x_{2,v}) + v_v$, with $x_{1,v}, x_{2,v}, v_v$ the three GINE-derived streams. For a local perturbation $\delta$ at $v$,
\begin{equation}
\frac{\partial u_v}{\partial \delta}
= \underbrace{\frac{\partial}{\partial \delta}\!\big[\sigma(x_{1,v}) \odot \mathrm{Long}(x_{2,v})\big]}_{\text{long-range (subject to smoothing)}}
+ \underbrace{\frac{\partial v_v}{\partial \delta}}_{\text{local bypass}}.
\end{equation}

The bypass \(v_v\) comes directly from 1-hop GINE and does not enter Chebyshev propagation, so local perturbation signals can remain available even when the long-range term becomes smoothed. Meanwhile, the gate \(\sigma(x_{1,v})\) adaptively regulates the contribution of the long-range path, preventing global context from uniformly overwhelming local distinctions.

%% file: tables/main_results.tex
\begin{table*}[t]
\centering
\caption{Aggregated performance over all 30 MoleculeACE targets. For each metric we report the mean (Avg.), mean rank, and number of targets where a method is the single best (\#Best, out of 30). Ranks and \#Best are computed among the methods listed; full per-target results are in Appendix Table~\ref{tab:full_per_target}. \textbf{Bold}: best, \underline{underline}: second-best. Lower Avg./Rank and higher \#Best are better.}
\label{tab:main_results}
\scalebox{1.00}{
\begin{tabular}{ll ccc ccc ccc}
\toprule
& & \multicolumn{3}{c}{\textbf{RMSE} ($\downarrow$)} & \multicolumn{3}{c}{$\textbf{RMSE}_{\textbf{cliff}}$ ($\downarrow$)} & \multicolumn{3}{c}{$\textbf{RMSE}_{\textbf{noncliff}}$ ($\downarrow$)} \\
\cmidrule(lr){3-5}\cmidrule(lr){6-8}\cmidrule(lr){9-11}
\textbf{Algorithm} & \textbf{Desc.} & \textbf{Avg.} & \textbf{Rank} & \textbf{\#Best} & \textbf{Avg.} & \textbf{Rank} & \textbf{\#Best} & \textbf{Avg.} & \textbf{Rank} & \textbf{\#Best} \\
\midrule
GraphCliff      & GRAPH  & \textbf{0.665} & \textbf{1.6} & \textbf{18} & \underline{0.757} & \underline{2.7} & \textbf{12} & \textbf{0.619} & \textbf{1.9} & \textbf{14} \\
SVM             & ECFP   & \underline{0.671} & \underline{2.0} & \underline{9} & \textbf{0.751} & \textbf{2.6} & \underline{10} & \underline{0.629} & \underline{2.2} & \underline{7} \\
GINE + PairNorm & GRAPH  & 0.706 & 4.1  & 2 & 0.799 & 5.0  & 1 & 0.662 & 4.0  & 6 \\
GINE + NodeNorm & GRAPH  & 0.717 & 4.6  & 0 & 0.797 & 4.9  & 0 & 0.680 & 5.0  & 0 \\
GINE + Residual & GRAPH  & 0.728 & 5.6  & 0 & 0.811 & 6.0  & 1 & 0.688 & 5.7  & 0 \\
Chemprop        & GRAPH  & 0.730 & 5.9  & 0 & 0.816 & 5.7  & 2 & 0.694 & 6.2  & 1 \\
LSTM            & SMILES & 0.742 & 6.9  & 1 & 0.871 & 9.2  & 0 & 0.678 & 5.5  & 2 \\
MLP             & ECFP   & 0.763 & 6.7  & 0 & 0.833 & 6.6  & 0 & 0.728 & 7.0  & 0 \\
GINE + DropEdge & GRAPH  & 0.794 & 9.1  & 0 & 0.878 & 9.0  & 0 & 0.756 & 8.8  & 0 \\
SCAGE (w/o 3D)  & GRAPH  & 0.800 & 8.8  & 0 & 0.813 & 5.7  & 3 & 0.789 & 9.6  & 0 \\
GCN             & GRAPH  & 0.914 & 11.3 & 0 & 0.974 & 11.2 & 0 & 0.892 & 11.0 & 0 \\
GAT             & GRAPH  & 0.947 & 12.0 & 0 & 1.005 & 12.1 & 0 & 0.925 & 12.0 & 0 \\
MPNN            & GRAPH  & 1.094 & 13.4 & 0 & 1.120 & 13.0 & 0 & 1.087 & 13.4 & 0 \\
$\mathrm{MolCLR}^{\text{pretrained}}_{\text{gcn}}$ & GRAPH & 1.218 & 13.9 & 0 & 1.234 & 13.3 & 0 & 1.213 & 14.0 & 0 \\
$\mathrm{MolCLR}^{\text{pretrained}}_{\text{gin}}$ & GRAPH & 1.220 & 14.3 & 0 & 1.217 & 13.8 & 0 & 1.223 & 14.4 & 0 \\
$\mathrm{ContextPred}^{\text{pretrained}}$     & GRAPH  & 1.629 & 16.1 & 0 & 1.768 & 16.1 & 0 & 1.530 & 15.7 & 0 \\
KPGT            & GRAPH  & 1.972 & 16.6 & 0 & 1.941 & 16.2 & 1 & 1.992 & 16.6 & 0 \\
\bottomrule
\end{tabular}}
\end{table*}

%% file: 3_results.tex
\section{Experimental Results}
\label{sec:results}
\subsection{Performance on MoleculeACE}

We evaluated our method on all 30 benchmark datasets provided by MoleculeACE. From the models reported in the original MoleculeACE study, we adopted six representative baselines spanning the three input modalities: SVM-ECFP, MLP-ECFP, LSTM-SMILES, GCN, GAT, and MPNN. We further included five models known for strong molecular property prediction: \textbf{ContextPred}~\citep{hu2019strategies}, a pretraining method predicting masked subgraphs from context; \textbf{MolCLR}~\citep{wang2022molecular}, a contrastive learning approach on molecular graphs; \textbf{Chemprop}~\citep{heid2023chemprop}, a directed MPNN; \textbf{KPGT}~\citep{li2023knowledge}, a knowledge-guided pretraining method; and \textbf{SCAGE}~\citep{qiao2025self}, a self-conformation-aware graph transformer (we excluded its 3D atom-distances for fair comparison with our 2D graph setting).

We also evaluate against anti-oversmoothing methods to examine whether architectures designed to mitigate representation collapse can better handle activity cliffs: GINE + PairNorm~\citep{zhao2019pairnorm}, GINE + NodeNorm~\citep{zhou2021understanding}, GINE + Residual, and GINE + DropEdge~\citep{rong2019dropedge}. We report three metrics: RMSE over all test molecules, measuring overall accuracy of predicted $-\log_{10} K_i$ values; $\mathrm{RMSE}_{\text{cliff}}$, computed only on activity cliff compounds to quantify error on these challenging, structure-sensitive samples; and $\mathrm{RMSE}_{\text{noncliff}}$, computed on the remaining non-cliff compounds.

Table~\ref{tab:main_results} aggregates performance over all 30 targets, reporting for each metric the mean value, mean rank, and the number of targets on which each method is the single best (\#Best). Complete per-target results are in Appendix Table~\ref{tab:full_per_target}. GraphCliff attains the best overall RMSE (mean 0.665, rank 1.6, best on 18 of 30 targets) and the best $\mathrm{RMSE}_{\text{noncliff}}$ (0.619, best on 14 targets), and on the more challenging $\mathrm{RMSE}_{\text{cliff}}$ metric it is the strongest learned model while remaining competitive with the best fingerprint method. We organize the comparison around four baseline families: ECFP-based machine learning, conventional message-passing GNNs, pretrained graph models, and anti-oversmoothing variants.

GraphCliff's closest competitor is SVM-ECFP. On cliff compounds the two are nearly tied. SVM-ECFP is marginally lower in mean $\mathrm{RMSE}_{\text{cliff}}$ (0.751 vs.\ 0.757), reflecting the strong local sensitivity of radius-based fingerprints that we analyze in Appendix Section~\ref{sec:error_analysis}. GraphCliff, however, is the single best on more targets (12 vs.\ 10) and surpasses SVM-ECFP on both overall RMSE (0.665 vs.\ 0.671) and $\mathrm{RMSE}_{\text{noncliff}}$ (0.619 vs.\ 0.629). Conventional message-passing GNNs fall far behind. GCN, GAT, and MPNN reach cliff RMSEs of 0.974, 1.005, and 1.120, more than $0.2$ above GraphCliff, consistent with the embedding collapse shown in Figure~\ref{fig:embedding_comparison}. 

Self-supervised pretraining offers no advantage here, as MolCLR and ContextPred rank at the bottom ($\mathrm{RMSE}_{\text{cliff}}$ $1.217$--$1.768$) because their objectives encode broad structural similarity rather than the localized changes that define cliffs. Anti-oversmoothing regularizers on a GINE backbone (PairNorm $0.799$, NodeNorm $0.797$, Residual $0.811$, DropEdge $0.878$) improve over plain GNNs but plateau well above GraphCliff, since they diversify node features uniformly instead of selectively modulating propagation at structurally sensitive atoms. Crucially, these cliff gains do not come at the cost of non-cliff accuracy, indicating a general representational improvement rather than a cliff-specific adjustment. Overall, \textbf{gated, node-wise control of propagation scale is more effective than uniform anti-oversmoothing techniques and existing graph architectures for activity cliff prediction}, yielding a single end-to-end graph model that is the strongest learned representation on cliff compounds while remaining best overall.

\subsection{Architecture Ablations}

We conducted ablation studies to assess the contributions of the short-range path, long-range path, and gating mechanism (Table~\ref{tab:ablation_pooling_combined}). Removing the short-range path caused the largest performance drop (RMSE $0.665 \to 1.288$, $+93.7\%$), underscoring its critical role. It captures one-hop message-passing information and feeds the long-range path, the gating mechanism, and the final sum fusion, so localized chemical information is preserved throughout the network. Removing the long-range path also degraded performance, but to a lesser extent ($0.665 \to 0.856$, $+28.7\%$), reflecting the smaller yet still substantial value of multi-hop context up to three hops. The gating mechanism had the smallest standalone effect ($0.665 \to 0.725$, $+9.0\%$) yet still contributed positively, proving more effective than naive feature summation at balancing local and global information. Notably, the two paths are strongly complementary. Each one alone is far weaker than their gated combination (short-only $1.001$, long-only $1.327$ RMSE), confirming that cliff prediction requires both local sensitivity and global context rather than either in isolation. For pooling, simple strategies exacerbate over-smoothing by uniformly aggregating indistinguishable node embeddings (Sum $0.963$, Mean $0.874$, Max $0.811$). SAGPool, which adaptively selects informative nodes by learned importance, achieves the lowest RMSE and $\mathrm{RMSE}_{\text{cliff}}$, confirming the benefit of adaptive node selection.

\input{tables/ablation}

\input{tables/graph_variant}

\begin{figure*}[htbp]
    \centering
    \includegraphics[width=0.95\textwidth]{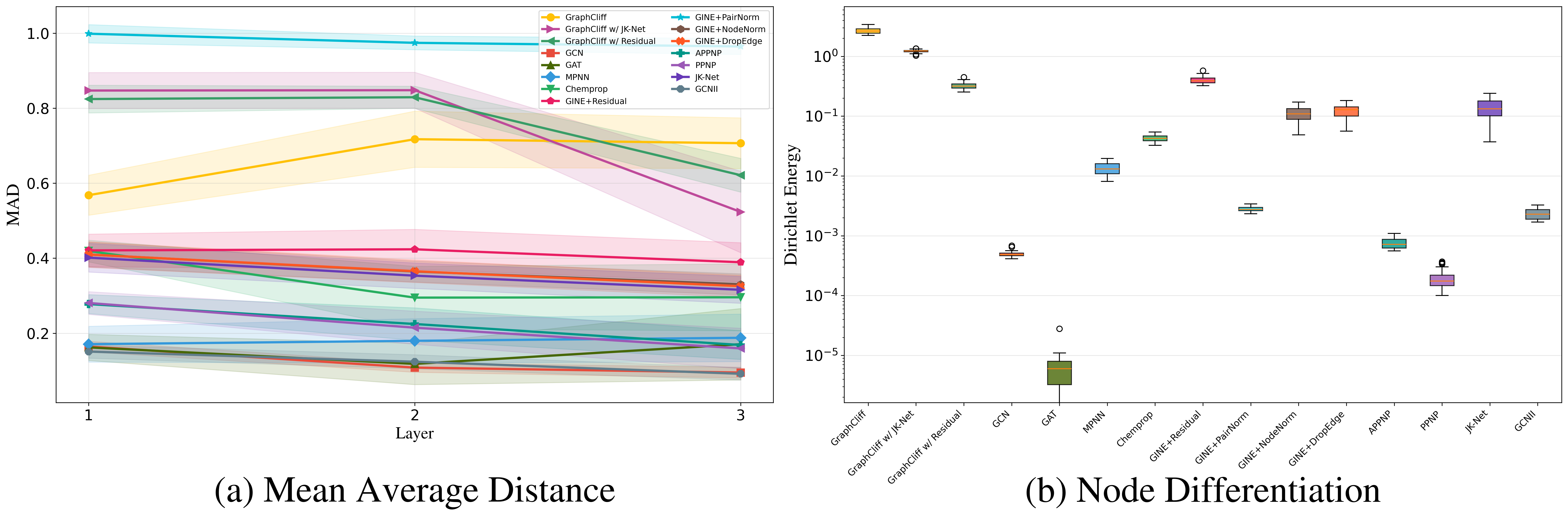}
    \caption{Layer-wise Mean Average Distance (MAD) and Dirichlet energy across GNN architectures. (a) Layer-wise MAD quantifies the degree of node-level differentiation preserved at each propagation depth. Higher MAD indicates better preservation of distinct node representations. (b) Dirichlet energy measures smoothness of node embeddings along edges. Lower energy indicates stronger smoothing.}
    \label{fig:comprehensive_analysis}
\end{figure*}

We next varied the GNN architectures used in the short- and long-range paths (Table~\ref{tab:graph_variant}). We first fix the long-range path to Chebyshev and replace the short-range GINE with GCN, GIN, or GAT (Table~\ref{tab:graph_variant}(a)). GINE incorporates edge features such as bond type, aromaticity, and conjugation, which is particularly beneficial for molecular graphs, whereas GCN and GIN do not explicitly use bond features. Accordingly, replacing GINE raises RMSE from $0.665$ to $0.724$ (GCN), $0.688$ (GIN), and $0.689$ (GAT), with $\mathrm{RMSE}_{\text{cliff}}$ rising correspondingly from $0.757$ to $0.819$, $0.777$, and $0.778$, confirming the benefit of edge-aware aggregation for molecular graphs. This sensitivity is especially relevant for activity cliffs, whose defining perturbation is often a single bond-level change that only edge-aware operators can register.

We then fix the short-range path to GINE and vary only the long-range path (Table~\ref{tab:graph_variant}(b)). Chebyshev outperforms stacked GNNs used for long-range propagation (GIN $0.694$, GAT $0.696$, GCN $0.715$ vs.\ $0.665$ RMSE), likely because spectral polynomials encode multi-hop information within a single layer rather than through repeated message passing. It also surpasses two schemes specifically designed to avoid deep stacking. APPNP applies an MLP once then propagates via Personalized PageRank. Since Chebyshev outperforms it ($0.665$ vs.\ $0.706$ RMSE; $0.757$ vs.\ $0.795$ $\mathrm{RMSE}_{\text{cliff}}$), the gain stems from Chebyshev's learnable polynomial basis combining multiple spectral components with task-specific coefficients, versus PageRank's fixed single-kernel diffusion. GCNII instead stacks many layers with initial-residual and identity mappings. Since Chebyshev also outperforms it ($0.665$ vs.\ $0.699$ RMSE; $0.757$ vs.\ $0.784$ $\mathrm{RMSE}_{\text{cliff}}$), residual tricks are insufficient. Thus Chebyshev's advantage is twofold: more expressive multi-hop propagation than fixed-kernel alternatives, and a single-layer design that preserves local detail better than deep stacking. 

%% file: tables/ablation.tex
\begin{table}[t]
\centering
\caption{Performance comparison across different module and pooling methods. The bold row is our default configuration (Short, Long, Gating, SAGPool (SAG)); $\Delta$ columns report the relative increase over it.}
\label{tab:ablation_pooling_combined}
\resizebox{\columnwidth}{!}{
\begin{tabular}{ccc|c|cc|cc}
\hline
\textbf{Short} & \textbf{Long} & \textbf{Gating} & \textbf{Pool} & \textbf{RMSE} & $\boldsymbol{\Delta}$ & $\textbf{RMSE}_{\textbf{cliff}}$ & $\boldsymbol{\Delta}_{\textbf{cliff}}$ \\
\hline
\textbf{O} & \textbf{O} & \textbf{O} & \textbf{SAG} & \textbf{0.665} & – & \textbf{0.757} & – \\
O & O & – & SAG & 0.725 & +9.0\% & 0.798 & +5.4\% \\
O & – & O & SAG & 0.856 & +28.7\% & 0.933 & +23.2\% \\
– & O & O & SAG & 1.288 & +93.7\% & 1.287 & +70.0\% \\
O & – & – & SAG & 1.001 & +50.5\% & 1.038 & +37.1\% \\
– & O & – & SAG & 1.327 & +99.5\% & 1.314 & +73.6\% \\
– & – & O & SAG & 1.361 & +104.7\% & 1.286 & +69.9\% \\
%– & – & – & SAG & 1.554 & +130.8\% & 1.481 & +93.4\% \\
\hline
O & O & O & Max     & 0.811 & +22.0\% & 0.871 & +15.1\% \\
O & O & O & Mean    & 0.874 & +31.4\% & 0.950 & +25.5\% \\
O & O & O & Sum     & 0.963 & +44.8\% & 1.024 & +35.3\% \\
\hline
\end{tabular}}
\end{table}
%78

%% file: tables/graph_variant.tex
\begin{table}[t]
\centering
\caption{Ablations for the short- and long-range paths over the 30 targets. \textbf{(a)} varies the short-range path with the long-range path fixed to Chebyshev; \textbf{(b)} varies the long-range path with the short-range path fixed to GINE. The bold row is our default configuration (GINE short-range, Chebyshev long-range); $\Delta$ columns report the relative increase over it.}
\label{tab:graph_variant}
\resizebox{\columnwidth}{!}{
\begin{tabular}{llcccc}
\toprule
\textbf{Short} & \textbf{Long} & \textbf{RMSE} & $\boldsymbol{\Delta}$ & $\textbf{RMSE}_{\textbf{cliff}}$ & $\boldsymbol{\Delta}_{\textbf{cliff}}$ \\
\midrule
\multicolumn{6}{l}{\textit{(a) Short-range path \,(long-range = Chebyshev)}} \\
\textbf{GINE} & \textbf{Chebyshev} & \textbf{0.665} & -- & \textbf{0.757} & -- \\
GCN  & Chebyshev & 0.724 & +8.9\% & 0.819 & +8.2\% \\
GIN  & Chebyshev & 0.688 & +3.5\% & 0.777 & +2.6\% \\
GAT  & Chebyshev & 0.689 & +3.6\% & 0.778 & +2.8\% \\

\midrule
\multicolumn{6}{l}{\textit{(b) Long-range path \,(short-range = GINE)}} \\
\textbf{GINE} & \textbf{Chebyshev} & \textbf{0.665} & -- & \textbf{0.757} & -- \\
GINE & GCNII & 0.699 & +5.1\% & 0.784 & +3.6\% \\
GINE & APPNP & 0.706 & +6.2\% & 0.795 & +5.0\% \\
GINE & GCN & 0.715 & +7.5\% & 0.803 & +6.1\% \\
GINE & GIN & 0.694 & +4.4\% & 0.774 & +2.2\% \\
GINE & GAT & 0.696 & +4.7\% & 0.778 & +2.8\% \\
\bottomrule
\end{tabular}
}
\end{table}

%% file: 4_analysis.tex
\section{Analysis}
\label{sec:analysis}

\begin{figure*}[t]
    \centering
    \includegraphics[width=0.95\textwidth]{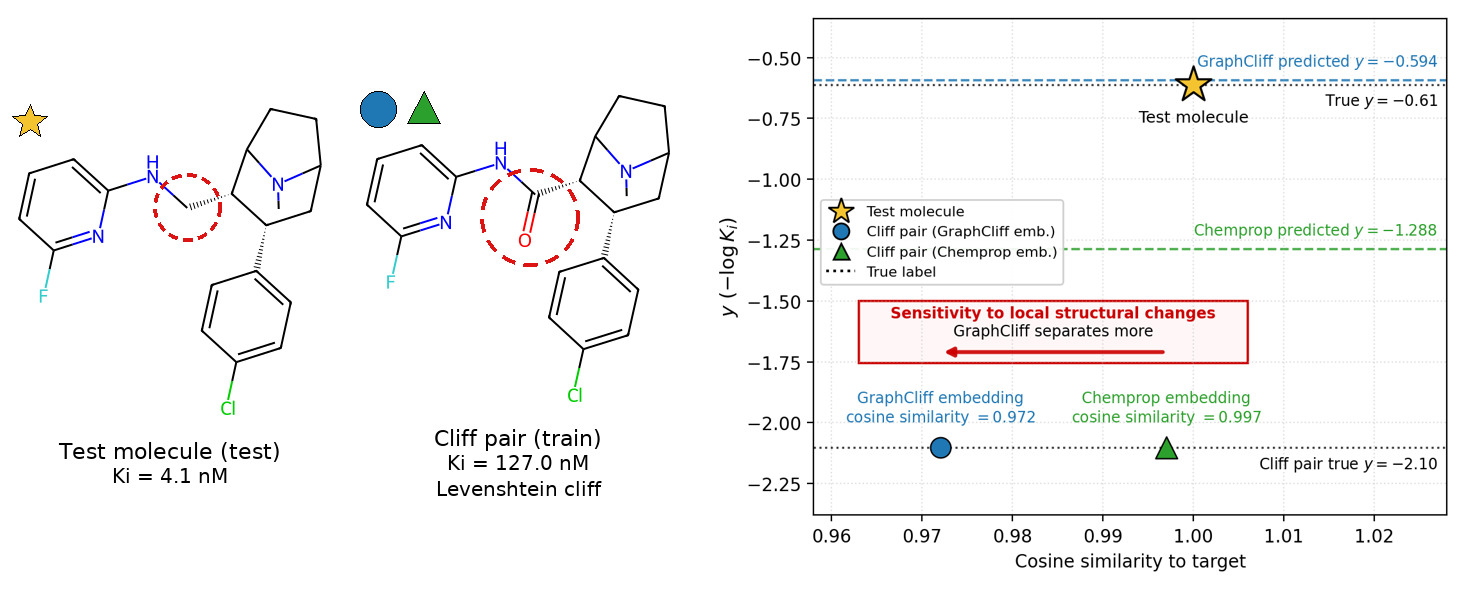}
    \caption{
    \textbf{Levenshtein activity cliff in CHEMBL238 (Ki).}
    The pair shows the target (test) and cliff (train) molecules with their embedding positions and model predictions. Chemprop maps the pair almost identically, whereas GraphCliff assigns a larger separation and provides a more accurate prediction.
    }
\label{fig:lev_cliff}
\end{figure*}

\subsection{Over-smoothing Mitigation}
\label{sec:oversmoothing_analysis}

Over-smoothing is a well-known limitation in deep GNNs, where repeated message passing causes node representations to become increasingly similar and lose their discriminative power~\citep{wu2023demystifying}. To quantify this phenomenon, we use two standard metrics: Mean Average Distance (MAD) and Dirichlet energy~\citep{chen2020measuring,hariri2025chebnet}. MAD measures how much node embeddings within a graph differ from one another at each layer. Specifically, we compute the cosine distance between every pair of node representations within a layer, exclude self-comparisons, and average these distances at the node level. Layer-wise MAD is then computed by averaging these node-level distances. A decrease in MAD with depth indicates that node representations are becoming more similar, signaling the onset of over-smoothing. We also evaluate the Dirichlet energy of the final-layer embeddings, which captures how much adjacent nodes differ in their representations. Low Dirichlet energy indicates that neighboring nodes have become nearly identical, whereas higher values suggest that important local variations are preserved.

As shown in Figure~\ref{fig:comprehensive_analysis} (a), most baseline models show a clear decrease in MAD as layers deepen. This progressive loss of discriminative power suggests that repeated propagation causes node representations to collapse, consistent with the $\mathcal{O}(|\lambda_2|^{L})$ decay predicted in Section~\ref{sec:theory}. Even models equipped with anti-oversmoothing mechanisms such as residual connections, DropEdge, and NodeNorm show limited effectiveness in deeper layers. In contrast, GraphCliff exhibits a markedly different trend. Its MAD either increases or remains consistently high across layers, suggesting that it preserves, and in some cases amplifies, node-level distinctions during multi-hop propagation. GINE+PairNorm also maintains relatively high MAD, as PairNorm explicitly re-centers and re-scales embeddings at each layer, preventing representation contraction.

The Dirichlet energy results point to the same conclusion (Figure~\ref{fig:comprehensive_analysis}(b)). Standard architectures such as GAT exhibit very low values, reaching $10^{-5}$, indicating that neighboring node embeddings become highly similar. Anti-oversmoothing variants achieve higher energy values but remain within a limited range, whereas GraphCliff attains the highest Dirichlet energy among all models, suggesting that it better preserves differences between adjacent nodes in the final representations. This preservation is consistent with its stronger cliff performance in Table~\ref{tab:main_results}. Overall, compared with models whose node-wise information becomes progressively less distinguishable with depth, GraphCliff maintains higher dispersion in both MAD and Dirichlet energy. This property is beneficial for activity cliff scenarios, where subtle structural changes should remain distinguishable in the learned representations.

\subsection{Distinguishing an Activity Cliff Pair}

MoleculeACE defines activity cliffs as compound pairs that are highly similar in structure (by substructure, scaffold, or SMILES similarity) yet differ at least 10-fold in \(K_i\). Such pairs are difficult for standard GNNs because a single atom- or bond-level change leaves most local neighborhoods almost unchanged, so message passing can map the two molecules to nearly identical embeddings even when their activities differ sharply.

Figure~\ref{fig:lev_cliff} examines one such pair from CHEMBL238 (Ki) in detail. The held-out test compound ($K_i = 4.1$\,nM, $y = -0.61$) and its Levenshtein-type training partner ($K_i = 127.0$\,nM, $y = -2.10$) are identical apart from a single edit at the linker, where a methylene ($\mathrm{CH_2}$) becomes a carbonyl ($\mathrm{C{=}O}$); despite this one-site change, potency drops roughly $31$-fold ($\Delta y \approx 1.49$ in $-\log_{10} K_i$), far above the $10$-fold cliff threshold. The two models treat this pair very differently. Chemprop maps the training molecule (green triangle) almost exactly onto the test target (star) with cosine similarity $0.997$, effectively collapsing the two analogues, whereas GraphCliff assigns a visibly lower similarity of $0.972$ (blue circle), separating them in proportion to the bond-level edit its short-range path registers. This propagates to prediction: against the true label $-0.6128$, GraphCliff predicts $-0.5936$ (error $0.02$) while Chemprop predicts $-1.2879$ (error $0.68$), pulled toward its near-identical training neighbor at $-2.10$ and inheriting that neighbor's much lower potency.

This case concretely illustrates the failure mode behind the aggregate results. Conventional GNN embeddings collapse distinct molecules into nearly identical representations precisely in the cliff regime, leaving the regression head with no signal that the two compounds differ, so it defaults to a neighbor-averaged prediction and misses the cliff. By preserving sensitivity to fine-grained structural perturbations through its local bypass, GraphCliff keeps such pairs apart and more faithfully models the abrupt activity changes that define the benchmark.

\subsection{Interpreting the Gating Mechanism}

To assess whether the proposed model captures functionally relevant substructures, we quantify how well the gate localizes cliff-defining atoms across the benchmark. Because each filter computes $u=\sigma(x_1)\odot\mathrm{Long}(x_2)+v$ with a residual connection, a small gate suppresses the long-range smoothing path and lets the local bypass dominate, so low-gate atoms are those that retain their local signal, exactly the behavior expected at cliff-defining sites. For every activity cliff pair we therefore rank atoms by their mean Layer-2 gate value $\sigma(x_1)$, take the $k$ with the smallest values, and compute their overlap with the atoms that structurally differ between the two compounds. Table~\ref{tab:topk_alignment} reports this overlap averaged over all 30 datasets, stratified by the split membership of each pair, alongside a random-selection baseline. The low-gate atoms align with cliff-defining atoms well above chance at every $k$ and across all three split categories (e.g.\ $60.4\%$ vs.\ $49.8\%$ at $k{=}5$ for train--train pairs), and the margin over random is preserved for test--test pairs, indicating that the alignment is not an artifact of memorized training structures but generalizes to held-out compounds. This provides quantitative support that the gating mechanism selectively preserves the substructures responsible for activity cliffs.

\begin{figure}[t]
    \centering
    \includegraphics[width=\columnwidth]{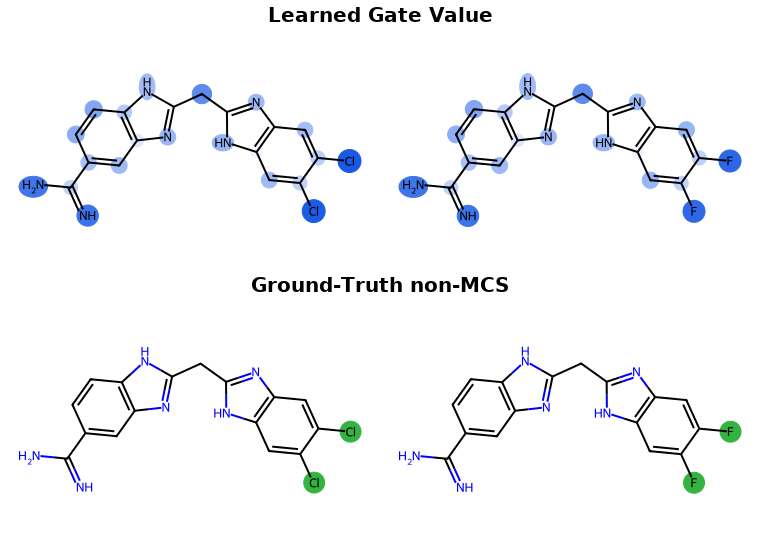}
    \caption{\textbf{Per-atom gating preserves the cliff-defining substructure.} A test--test cliff pair from CHEMBL244 (Ki) differing only in a benzimidazole dihalogen (5,6-dichloro, $K_i=140$\,nM vs.\ 5,6-difluoro, $K_i=6$\,nM; $\sim$23-fold). \textbf{Top:} atoms by mean Layer-2 gate $\sigma(x_1)$ (blue\,=\,low gate, white\,=\,high); \textbf{Bottom:} non-MCS atoms (green). The two cliff-defining halogens are the two lowest-gate atoms, consistent with Table~\ref{tab:topk_alignment}.}
    \label{fig:gate_heatmap}
\end{figure}

To illustrate this behavior on a concrete pair, we visualize the atom-level gate values from the sigmoid-gated vector \( \sigma(x_1) \in \mathbb{R}^{N \times d} \) for a single held-out activity cliff pair. Figure~\ref{fig:gate_heatmap} presents this pair. The bottom row highlights the atoms that differ between the two compounds, while the top row shows the corresponding mean atom-level gate values (darker blue indicating \emph{lower} gate, i.e.\ stronger local bypass). The low-gate atoms align with the localized structural differences that induce the large activity change, indicating that the gating path selectively preserves chemically meaningful perturbations rather than admitting global context everywhere, exactly the behavior needed for cliff compounds where small, localized modifications cause disproportionate functional effects.

\input{tables/gating}

%% file: tables/gating.tex
\begin{table}[t]
\centering
\caption{\textbf{Top-$k$ gate alignment.} Since each filter computes $u=\sigma(x_1)\odot\mathrm{Long}(x_2)+v$, low-gate atoms keep their local signal, as expected at cliff sites. For each cliff pair we take the $k$ atoms with the smallest mean Layer-2 gate $\sigma(x_1)$ and report their overlap (\%) with the non-MCS atoms, averaged over 30 datasets and stratified by split; \emph{Random} selects $k$ atoms at random. Low-gate atoms align with cliff-defining atoms well above chance at every $k$.}
\label{tab:topk_alignment}
\scalebox{0.95}{
\begin{tabular}{l cc cc cc}
\toprule
& \multicolumn{2}{c}{\textbf{Top-1}} & \multicolumn{2}{c}{\textbf{Top-3}} & \multicolumn{2}{c}{\textbf{Top-5}} \\
\cmidrule(lr){2-3}\cmidrule(lr){4-5}\cmidrule(lr){6-7}
\textbf{Cliff-pair} & \textbf{Gate} & Random & \textbf{Gate} & Random & \textbf{Gate} & Random \\
\midrule
train--train & \textbf{23.9} & 15.5 & \textbf{45.0} & 36.2 & \textbf{60.4} & 49.8 \\
train--test  & \textbf{23.0} & 15.0 & \textbf{43.9} & 35.5 & \textbf{58.8} & 49.0 \\
test--test   & \textbf{22.1} & 14.2 & \textbf{42.9} & 33.7 & \textbf{57.6} & 46.7 \\
\bottomrule
\end{tabular}}
\end{table}

%% file: 5_conclusion.tex
\section{Limitations}
On a subset of the 30 targets, SVM-ECFP outperforms GraphCliff on cliff pairs, where the strong local sensitivity of radius-based fingerprints yields lower RMSE. This edge persists even though GraphCliff attains the best overall RMSE and SVM-ECFP ranks only second on average. Although GraphCliff is designed to enhance local sensitivity by integrating short- and long-range representations, its learned graph embedding can reduce separation among globally similar molecules, weakening discrimination for fine-grained perturbations. A more detailed error analysis of these failure cases is provided in Appendix Section~\ref{sec:error_analysis}, where we argue that the residual gap is structurally inherent to the contrast between fixed binary fingerprints and learned continuous embeddings, and identify two concrete remedies: cliff-aware training objectives (e.g.\ contrastive or margin-based losses that explicitly separate cliff pairs) and hybrid representations that inject ECFP features into the graph model. While GraphCliff improves robustness relative to conventional GNNs, capturing extremely localized structure–activity relationships remains challenging, and we view closing the remaining gap through such cliff-aware training as the most promising next step.

\section{Conclusion}
We introduced GraphCliff, a graph architecture for activity cliff prediction that resolves the tension between local perturbation sensitivity and global context. Rather than mixing the two uniformly, GraphCliff routes a short-range (GINE) signal and a long-range (Chebyshev) signal through a node-level, locally conditioned gate together with a local bypass, letting each atom decide how much global context to admit. Across all 30 MoleculeACE targets, this yields the best overall and non-cliff RMSE and remains competitive with the strongest fingerprint method on cliff compounds, while decisively outperforming both conventional GNNs and uniform anti-oversmoothing techniques. Our analyses explain why. GraphCliff preserves node-level distinctions where standard GNNs collapse them (higher MAD and Dirichlet energy), and its gate selectively attends to the substructures that define each cliff. These results indicate that \emph{node-wise, adaptive control of propagation scale}, not uniform regularization, is the key to making graph models cliff-aware. The remaining gap to fixed fingerprints is structurally inherent rather than a tuning artifact. Closing it through cliff-aware training objectives or hybrid ECFP features is a promising direction.

%% file: tables/full_results.tex
\begin{table}[H]
\centering
\scriptsize
\setlength{\tabcolsep}{4pt}\renewcommand{\arraystretch}{0.78}
\caption{Per-target RMSE and $\text{RMSE}_{\text{cliff}}$ ($\downarrow$) for all methods. Best per target in \textbf{bold}, second-best \underline{underlined}.}
\label{tab:full_per_target}
\textbf{(a) RMSE}\\[1pt]
\resizebox{\linewidth}{!}{
\begin{tabular}{lccccccccccccccccccccc}
\toprule
\textbf{Target} & \rotatebox{90}{GraphCliff} & \rotatebox{90}{SVM (ECFP)} & \rotatebox{90}{GINE+PairNorm} & \rotatebox{90}{GINE+NodeNorm} & \rotatebox{90}{GINE+Residual} & \rotatebox{90}{Chemprop} & \rotatebox{90}{LSTM} & \rotatebox{90}{MLP (ECFP)} & \rotatebox{90}{GINE+DropEdge} & \rotatebox{90}{SCAGE} & \rotatebox{90}{GCN} & \rotatebox{90}{GAT} & \rotatebox{90}{MPNN} & \rotatebox{90}{MolCLR$^{\text{pre}}_{\text{gcn}}$} & \rotatebox{90}{MolCLR$^{\text{pre}}_{\text{gin}}$} & \rotatebox{90}{ContextPred$^{\text{pre}}$} & \rotatebox{90}{KPGT} & \rotatebox{90}{APPNP} & \rotatebox{90}{PPNP} & \rotatebox{90}{JK-Net} & \rotatebox{90}{GCNII} \\
\midrule
1862 (Ki) & 0.781 & \underline{0.774} & 0.841 & 0.921 & 0.866 & 0.815 & \textbf{0.761} & 0.878 & 1.006 & 0.875 & 0.942 & 0.987 & 0.948 & 1.068 & 1.051 & 1.351 & 1.668 & 1.008 & 1.463 & 0.972 & 0.960 \\
1871 (Ki) & \textbf{0.628} & 0.665 & \underline{0.630} & 0.672 & 0.678 & 0.704 & 0.662 & 0.737 & 0.770 & 0.758 & 0.769 & 0.798 & 1.058 & 0.925 & 1.077 & 1.714 & 1.976 & 0.762 & 1.140 & 0.640 & 0.740 \\
2034 (Ki) & 0.747 & \textbf{0.674} & 0.725 & 0.719 & 0.730 & 0.775 & 0.755 & 0.742 & 0.749 & 0.745 & 0.810 & 0.809 & 0.905 & 1.292 & 1.510 & 1.415 & 1.420 & 0.765 & 1.004 & \underline{0.714} & 0.759 \\
2047 (EC50) & \underline{0.592} & 0.614 & 0.613 & 0.654 & 0.670 & 0.693 & 0.696 & 0.677 & 0.742 & 0.714 & 0.797 & 0.840 & 1.030 & 0.767 & 0.784 & 1.654 & 2.837 & 0.759 & 1.256 & \textbf{0.592} & 0.725 \\
204 (Ki) & \textbf{0.691} & \underline{0.723} & 0.745 & 0.738 & 0.800 & 0.811 & 0.822 & 0.815 & 0.866 & 0.814 & 1.056 & 1.138 & 1.458 & 1.557 & 1.689 & 1.957 & 2.302 & 0.872 & 1.524 & 0.778 & 0.826 \\
2147 (Ki) & \textbf{0.560} & \underline{0.576} & 0.693 & 0.779 & 0.711 & 0.649 & 0.646 & 0.723 & 0.785 & 0.910 & 0.840 & 0.966 & 1.024 & 1.905 & 1.226 & 1.266 & 1.675 & 1.107 & 1.977 & 0.666 & 1.047 \\
214 (Ki) & \textbf{0.621} & \underline{0.634} & 0.722 & 0.681 & 0.710 & 0.660 & 0.723 & 0.693 & 0.764 & 0.779 & 1.007 & 1.051 & 1.183 & 1.050 & 1.144 & 1.501 & 1.802 & 0.800 & 1.134 & 0.666 & 0.754 \\
218 (EC50) & \underline{0.697} & 0.719 & \textbf{0.690} & 0.765 & 0.727 & 0.758 & 0.748 & 0.785 & 0.852 & 0.738 & 0.928 & 0.957 & 1.053 & 1.072 & 1.100 & 1.726 & 2.199 & 0.853 & 1.052 & 0.735 & 0.758 \\
219 (Ki) & \textbf{0.664} & 0.710 & 0.720 & 0.702 & 0.747 & \underline{0.692} & 0.780 & 0.756 & 0.833 & 0.876 & 1.026 & 0.979 & 0.903 & 0.984 & 1.022 & 1.661 & 2.121 & 0.886 & 1.071 & 0.740 & 0.816 \\
228 (Ki) & \textbf{0.651} & \underline{0.662} & 0.693 & 0.702 & 0.708 & 0.670 & 0.779 & 0.755 & 0.810 & 0.771 & 0.958 & 1.026 & 1.000 & 1.340 & 1.459 & 1.494 & 1.856 & 0.844 & 1.196 & 0.713 & 0.784 \\
231 (Ki) & \textbf{0.708} & 0.750 & \underline{0.729} & 0.757 & 0.795 & 0.760 & 0.809 & 1.333 & 0.792 & 0.914 & 0.878 & 0.991 & 1.305 & 1.395 & 1.643 & 1.975 & 2.605 & 0.866 & 1.320 & 0.776 & 0.857 \\
233 (Ki) & 0.786 & \textbf{0.774} & 0.796 & 0.791 & 0.794 & 0.799 & 0.850 & 0.845 & 0.861 & 0.824 & 1.056 & 1.066 & 1.074 & 1.229 & 1.299 & 1.811 & 1.670 & 0.904 & 1.297 & \underline{0.777} & 0.837 \\
234 (Ki) & \textbf{0.618} & \underline{0.622} & 0.682 & 0.672 & 0.694 & 0.687 & 0.738 & 0.669 & 0.752 & 0.799 & 0.934 & 0.950 & 0.959 & 1.325 & 1.347 & 1.349 & 1.676 & 0.794 & 1.135 & 0.655 & 0.729 \\
235 (EC50) & 0.650 & \textbf{0.640} & \underline{0.647} & 0.656 & 0.688 & 0.708 & 0.727 & 0.718 & 0.712 & 0.694 & 0.901 & 0.869 & 1.058 & 1.028 & 1.313 & 1.615 & 2.670 & 0.761 & 1.067 & 0.672 & 0.693 \\
236 (Ki) & \textbf{0.697} & \underline{0.698} & 0.757 & 0.730 & 0.736 & 0.799 & 0.812 & 0.733 & 0.814 & 0.799 & 0.943 & 1.002 & 1.364 & 1.315 & 1.414 & 1.483 & 2.154 & 0.862 & 1.332 & 0.741 & 0.786 \\
237 (EC50) & \textbf{0.708} & \underline{0.720} & 0.757 & 0.738 & 0.771 & 0.767 & 0.783 & 0.902 & 0.895 & 0.967 & 1.132 & 1.103 & 1.402 & 1.209 & 1.089 & 1.665 & 1.691 & 0.929 & 1.443 & 0.764 & 1.091 \\
237 (Ki) & \underline{0.705} & \textbf{0.677} & 0.738 & 0.728 & 0.753 & 0.743 & 0.773 & 0.722 & 0.782 & 0.762 & 1.112 & 1.085 & 1.053 & 1.338 & 1.345 & 1.564 & 2.158 & 0.829 & 1.340 & 0.747 & 0.769 \\
238 (Ki) & \textbf{0.597} & \underline{0.610} & 0.658 & 0.666 & 0.703 & 0.673 & 0.654 & 0.684 & 0.744 & 0.726 & 0.937 & 0.928 & 1.142 & 1.168 & 1.198 & 1.609 & 2.389 & 0.785 & 1.152 & 0.644 & 0.727 \\
239 (EC50) & \textbf{0.670} & \underline{0.678} & 0.733 & 0.693 & 0.710 & 0.819 & 0.765 & 0.756 & 0.792 & 0.796 & 0.905 & 0.902 & 1.288 & 0.944 & 1.054 & 1.836 & 2.751 & 0.823 & 1.098 & 0.786 & 0.756 \\
244 (Ki) & \textbf{0.668} & \underline{0.715} & 0.786 & 0.758 & 0.765 & 0.726 & 0.800 & 0.795 & 0.847 & 0.892 & 1.075 & 1.088 & 1.659 & 1.897 & 1.849 & 1.894 & 2.126 & 0.990 & 1.604 & 0.758 & 0.857 \\
262 (Ki) & \underline{0.752} & \textbf{0.724} & 0.789 & 0.807 & 0.859 & 0.868 & 0.767 & 0.904 & 0.919 & 0.923 & 0.934 & 0.994 & 1.021 & 1.220 & 1.174 & 1.981 & 2.534 & 0.980 & 1.092 & 0.837 & 0.912 \\
264 (Ki) & \underline{0.619} & \textbf{0.615} & 0.643 & 0.658 & 0.663 & 0.637 & 0.665 & 0.672 & 0.744 & 0.705 & 0.855 & 0.896 & 1.082 & 1.067 & 1.081 & 1.661 & 1.514 & 0.794 & 1.077 & 0.658 & 0.704 \\
2835 (Ki) & \underline{0.396} & 0.420 & \textbf{0.384} & 0.415 & 0.429 & 0.433 & 0.431 & 0.488 & 0.478 & 0.505 & 0.505 & 0.555 & 0.668 & 1.343 & 1.066 & 1.542 & 0.607 & 0.502 & 0.977 & 0.421 & 0.476 \\
287 (Ki) & \underline{0.706} & 0.714 & 0.734 & 0.733 & 0.742 & 0.709 & 0.791 & 0.733 & 0.833 & 0.801 & 0.886 & 0.947 & 0.927 & 0.873 & 0.873 & 1.516 & 1.473 & 0.794 & 1.042 & \textbf{0.702} & 0.804 \\
2971 (Ki) & \underline{0.615} & \textbf{0.605} & 0.683 & 0.647 & 0.692 & 0.745 & 0.689 & 0.674 & 0.679 & 0.745 & 0.781 & 0.803 & 0.973 & 1.818 & 1.232 & 1.764 & 1.337 & 0.840 & 1.384 & 0.654 & 0.860 \\
3979 (EC50) & \textbf{0.623} & \underline{0.629} & 0.631 & 0.689 & 0.678 & 0.711 & 0.740 & 0.661 & 0.752 & 0.921 & 0.812 & 0.923 & 1.183 & 1.053 & 0.987 & 1.623 & 2.142 & 0.822 & 1.136 & 0.684 & 0.764 \\
4005 (Ki) & \textbf{0.617} & 0.646 & 0.667 & 0.674 & 0.650 & 0.709 & 0.764 & 0.680 & 0.704 & 0.705 & 0.875 & 0.861 & 0.998 & 1.045 & 1.121 & 1.557 & 1.506 & 0.721 & 1.079 & \underline{0.622} & 0.806 \\
4203 (Ki) & \underline{0.900} & \textbf{0.880} & 0.931 & 0.956 & 0.940 & 1.003 & 0.907 & 0.947 & 0.950 & 1.024 & 0.975 & 1.004 & 1.056 & 1.073 & 1.058 & 1.737 & 2.536 & 1.006 & 1.069 & 0.977 & 0.977 \\
4616 (EC50) & \textbf{0.634} & \underline{0.635} & 0.692 & 0.675 & 0.660 & 0.704 & 0.739 & 0.727 & 0.748 & 0.740 & 0.867 & 0.873 & 0.934 & 0.899 & 0.909 & 1.238 & 1.474 & 0.797 & 0.929 & 0.714 & 0.738 \\
4792 (Ki) & \underline{0.635} & \textbf{0.633} & 0.676 & 0.750 & 0.762 & 0.675 & 0.691 & 0.692 & 0.837 & 0.781 & 0.923 & 1.004 & 1.122 & 1.348 & 1.498 & 1.706 & 2.304 & 0.811 & 1.176 & 0.680 & 0.792 \\
\bottomrule
\end{tabular}}

\vspace{8pt}

\textbf{(b) RMSE$_{\text{cliff}}$}\\[1pt]
\resizebox{\linewidth}{!}{
\begin{tabular}{lccccccccccccccccccccc}
\toprule
\textbf{Target} & \rotatebox{90}{GraphCliff} & \rotatebox{90}{SVM (ECFP)} & \rotatebox{90}{GINE+PairNorm} & \rotatebox{90}{GINE+NodeNorm} & \rotatebox{90}{GINE+Residual} & \rotatebox{90}{Chemprop} & \rotatebox{90}{LSTM} & \rotatebox{90}{MLP (ECFP)} & \rotatebox{90}{GINE+DropEdge} & \rotatebox{90}{SCAGE} & \rotatebox{90}{GCN} & \rotatebox{90}{GAT} & \rotatebox{90}{MPNN} & \rotatebox{90}{MolCLR$^{\text{pre}}_{\text{gcn}}$} & \rotatebox{90}{MolCLR$^{\text{pre}}_{\text{gin}}$} & \rotatebox{90}{ContextPred$^{\text{pre}}$} & \rotatebox{90}{KPGT} & \rotatebox{90}{APPNP} & \rotatebox{90}{PPNP} & \rotatebox{90}{JK-Net} & \rotatebox{90}{GCNII} \\
\midrule
1862 (Ki) & \underline{0.674} & \textbf{0.674} & 0.746 & 0.856 & 0.785 & 0.693 & 0.793 & 0.781 & 0.999 & 0.677 & 0.942 & 1.001 & 0.888 & 1.194 & 1.176 & 1.813 & 1.465 & 0.948 & 1.437 & 0.961 & 0.919 \\
1871 (Ki) & \underline{0.797} & 0.873 & 0.867 & 0.834 & \textbf{0.781} & 0.919 & 0.850 & 0.958 & 0.954 & 0.823 & 1.009 & 1.042 & 1.154 & 0.832 & 1.020 & 1.766 & 1.822 & 0.969 & 1.333 & 0.823 & 0.938 \\
2034 (Ki) & 0.858 & \textbf{0.813} & 0.845 & 0.852 & 0.896 & 0.886 & 0.944 & 0.849 & 0.901 & 0.825 & 0.928 & 0.942 & 0.927 & 1.305 & 1.550 & 1.412 & 1.594 & 0.900 & 0.931 & \underline{0.825} & 0.857 \\
2047 (EC50) & 0.655 & 0.687 & \underline{0.595} & 0.720 & 0.673 & 0.720 & 0.790 & 0.728 & 0.730 & 0.605 & 0.781 & 0.790 & 0.951 & 0.739 & 0.764 & 2.007 & 2.695 & 0.771 & 0.928 & \textbf{0.578} & 0.700 \\
204 (Ki) & \textbf{0.821} & 0.859 & 0.898 & 0.864 & 0.938 & \underline{0.851} & 0.930 & 0.962 & 0.978 & 0.888 & 1.201 & 1.282 & 1.581 & 1.579 & 1.709 & 2.202 & 2.210 & 0.973 & 1.709 & 0.900 & 0.934 \\
2147 (Ki) & \textbf{0.579} & \underline{0.580} & 0.679 & 0.696 & 0.733 & 0.639 & 0.725 & 0.704 & 0.802 & 0.858 & 0.825 & 0.917 & 0.934 & 2.063 & 1.015 & 1.872 & 1.201 & 1.005 & 1.881 & 0.681 & 0.971 \\
214 (Ki) & \underline{0.726} & \textbf{0.724} & 0.809 & 0.793 & 0.811 & 0.825 & 0.851 & 0.780 & 0.874 & 0.834 & 1.084 & 1.137 & 1.243 & 1.076 & 1.183 & 1.593 & 1.780 & 0.889 & 1.152 & 0.768 & 0.834 \\
218 (EC50) & 0.781 & \textbf{0.761} & \underline{0.771} & 0.776 & 0.793 & 0.794 & 0.818 & 0.806 & 0.875 & 0.783 & 0.952 & 0.984 & 1.062 & 1.118 & 1.106 & 1.947 & 2.099 & 0.886 & 0.999 & 0.782 & 0.833 \\
219 (Ki) & \textbf{0.744} & 0.788 & \underline{0.769} & 0.797 & 0.819 & 0.774 & 0.855 & 0.832 & 0.894 & 0.863 & 1.055 & 0.982 & 0.919 & 0.940 & 0.991 & 1.695 & 2.099 & 0.924 & 1.066 & 0.813 & 0.861 \\
228 (Ki) & \textbf{0.674} & \underline{0.676} & 0.714 & 0.756 & 0.793 & 0.695 & 0.884 & 0.757 & 0.858 & 0.786 & 1.000 & 1.028 & 1.014 & 1.329 & 1.364 & 1.657 & 1.847 & 0.879 & 1.160 & 0.771 & 0.828 \\
231 (Ki) & 0.866 & 0.932 & 0.785 & 0.844 & 0.803 & 0.817 & 1.070 & 1.272 & 0.872 & \underline{0.784} & 0.797 & 0.970 & 1.223 & 1.420 & 1.659 & 2.135 & 2.622 & 0.878 & 1.226 & 0.858 & \textbf{0.778} \\
233 (Ki) & 0.880 & 0.859 & 0.890 & 0.877 & 0.881 & \underline{0.844} & 0.942 & 0.916 & 0.941 & \textbf{0.800} & 1.106 & 1.099 & 1.138 & 1.256 & 1.360 & 1.908 & 1.656 & 0.958 & 1.303 & 0.859 & 0.912 \\
234 (Ki) & \textbf{0.632} & \underline{0.632} & 0.703 & 0.654 & 0.700 & 0.653 & 0.797 & 0.676 & 0.742 & 0.811 & 0.919 & 0.912 & 0.922 & 1.335 & 1.357 & 1.489 & 1.708 & 0.779 & 1.060 & 0.656 & 0.710 \\
235 (EC50) & \textbf{0.754} & \underline{0.774} & 0.795 & 0.779 & 0.819 & 0.799 & 0.847 & 0.818 & 0.854 & 0.787 & 1.039 & 1.012 & 1.194 & 1.045 & 1.391 & 1.862 & 2.636 & 0.905 & 1.146 & 0.785 & 0.832 \\
236 (Ki) & \textbf{0.785} & \underline{0.798} & 0.817 & 0.817 & 0.798 & 0.883 & 0.905 & 0.810 & 0.910 & 0.928 & 1.000 & 1.100 & 1.454 & 1.296 & 1.382 & 1.713 & 2.057 & 0.916 & 1.321 & 0.833 & 0.830 \\
237 (EC50) & \textbf{0.767} & \underline{0.782} & 0.874 & 0.817 & 0.854 & 0.839 & 0.903 & 0.950 & 0.949 & 0.960 & 1.094 & 1.064 & 1.334 & 1.252 & 1.103 & 1.766 & 1.762 & 0.932 & 1.349 & 0.865 & 1.040 \\
237 (Ki) & 0.783 & \textbf{0.735} & 0.810 & 0.795 & 0.810 & 0.800 & 0.862 & \underline{0.765} & 0.844 & 0.822 & 1.152 & 1.098 & 1.109 & 1.365 & 1.390 & 1.710 & 2.094 & 0.866 & 1.421 & 0.824 & 0.831 \\
238 (Ki) & 0.729 & \textbf{0.681} & 0.736 & 0.747 & 0.765 & 0.736 & 0.792 & 0.732 & 0.831 & \underline{0.681} & 0.925 & 0.946 & 1.208 & 1.307 & 1.336 & 2.037 & 2.240 & 0.867 & 1.181 & 0.739 & 0.744 \\
239 (EC50) & \textbf{0.792} & \underline{0.819} & 0.864 & 0.823 & 0.839 & 0.827 & 0.905 & 0.901 & 0.930 & 0.851 & 1.024 & 1.012 & 1.482 & 0.997 & 1.112 & 2.099 & 2.660 & 0.948 & 1.179 & 0.911 & 0.867 \\
244 (Ki) & \textbf{0.752} & \underline{0.797} & 0.859 & 0.836 & 0.815 & 0.797 & 0.913 & 0.850 & 0.910 & 0.972 & 1.060 & 1.117 & 1.557 & 1.930 & 1.837 & 2.011 & 2.103 & 0.990 & 1.539 & 0.834 & 0.935 \\
262 (Ki) & \underline{0.702} & \textbf{0.656} & 0.935 & 0.804 & 0.879 & 1.028 & 0.781 & 0.948 & 0.875 & 0.845 & 1.004 & 1.032 & 1.036 & 1.090 & 1.066 & 2.061 & 2.663 & 0.954 & 1.161 & 0.861 & 0.938 \\
264 (Ki) & \underline{0.671} & 0.674 & 0.724 & 0.716 & 0.750 & \textbf{0.652} & 0.767 & 0.731 & 0.790 & 0.738 & 0.910 & 0.936 & 1.012 & 1.030 & 1.025 & 1.561 & 1.589 & 0.844 & 1.015 & 0.719 & 0.775 \\
2835 (Ki) & 0.795 & 0.743 & 0.750 & 0.756 & 0.824 & 0.762 & 0.840 & 0.876 & 0.917 & \underline{0.694} & 0.926 & 0.924 & 1.067 & 1.182 & 0.902 & 1.025 & \textbf{0.682} & 0.979 & 1.326 & 0.780 & 0.936 \\
287 (Ki) & 0.798 & 0.811 & 0.746 & 0.766 & 0.757 & \textbf{0.715} & 0.894 & 0.852 & 0.873 & 0.813 & 0.900 & 0.994 & 0.973 & 0.850 & 0.850 & 1.572 & 1.556 & 0.864 & 1.117 & \underline{0.727} & 0.866 \\
2971 (Ki) & 0.778 & \textbf{0.659} & 0.851 & 0.770 & 0.799 & 0.954 & 0.886 & 0.764 & 0.831 & \underline{0.726} & 0.917 & 0.966 & 0.945 & 1.840 & 1.102 & 1.627 & 1.428 & 0.853 & 1.305 & 0.789 & 0.885 \\
3979 (EC50) & \textbf{0.654} & \underline{0.674} & 0.690 & 0.703 & 0.715 & 0.770 & 0.790 & 0.724 & 0.788 & 0.894 & 0.805 & 0.914 & 1.145 & 0.921 & 0.828 & 1.867 & 2.054 & 0.796 & 1.003 & 0.715 & 0.799 \\
4005 (Ki) & \textbf{0.712} & 0.742 & 0.767 & 0.783 & \underline{0.741} & 0.808 & 0.900 & 0.769 & 0.801 & 0.770 & 0.909 & 0.901 & 1.016 & 1.043 & 1.132 & 1.494 & 1.590 & 0.831 & 1.088 & 0.773 & 0.850 \\
4203 (Ki) & 1.177 & \textbf{1.001} & 1.228 & 1.230 & 1.269 & 1.484 & 1.318 & 1.026 & 1.196 & \underline{1.016} & 1.211 & 1.208 & 1.149 & 1.502 & 1.487 & 2.027 & 2.640 & 1.203 & 1.099 & 1.094 & 1.150 \\
4616 (EC50) & 0.719 & \underline{0.692} & 0.770 & 0.700 & 0.710 & 0.795 & 0.831 & 0.778 & 0.761 & \textbf{0.686} & 0.831 & 0.835 & 0.860 & 0.816 & 0.824 & 1.332 & 1.468 & 0.773 & 0.842 & 0.750 & 0.740 \\
4792 (Ki) & \underline{0.651} & \textbf{0.638} & 0.697 & 0.734 & 0.771 & 0.713 & 0.750 & 0.682 & 0.861 & 0.859 & 0.926 & 1.014 & 1.114 & 1.356 & 1.499 & 1.783 & 2.211 & 0.813 & 1.169 & 0.695 & 0.800 \\
\bottomrule
\end{tabular}}
\end{table}

%% file: 6_appendix.tex
%% The wide results table is included (one-column) from main.tex on the
%% previous page. Switch back to two columns HERE (the column switch forces a
%% page break, which is the natural page-11 -> page-12 boundary). Both figures
%% span the full width as figure*, and the prose flows in two columns beneath
%% them on the SAME page. A column switch between the figures and the prose
%% would instead force the prose onto a new page, so there is none.
%% Put the full-width figures in the OPTIONAL argument of \twocolumn. This
%% material is typeset full width at the very top of the new two-column page as
%% ordinary (non-floating) content, so it cannot be deferred to the next page
%% the way a figure* float is. The error-analysis prose then flows in two
%% columns directly beneath it on the SAME page.
%% \subfloat breaks outside a float environment, so use subcaption's
%% \subcaptionbox (works anywhere) and declare \captionsetup{type=figure} so
%% \captionof numbers the figures. Trailing %% remove inter-panel spaces that
%% would otherwise overflow the line.
\twocolumn[%
    \begin{center}
    \captionsetup{type=figure}%
    \subcaptionbox{GraphCliff}{\includegraphics[width=0.31\linewidth]{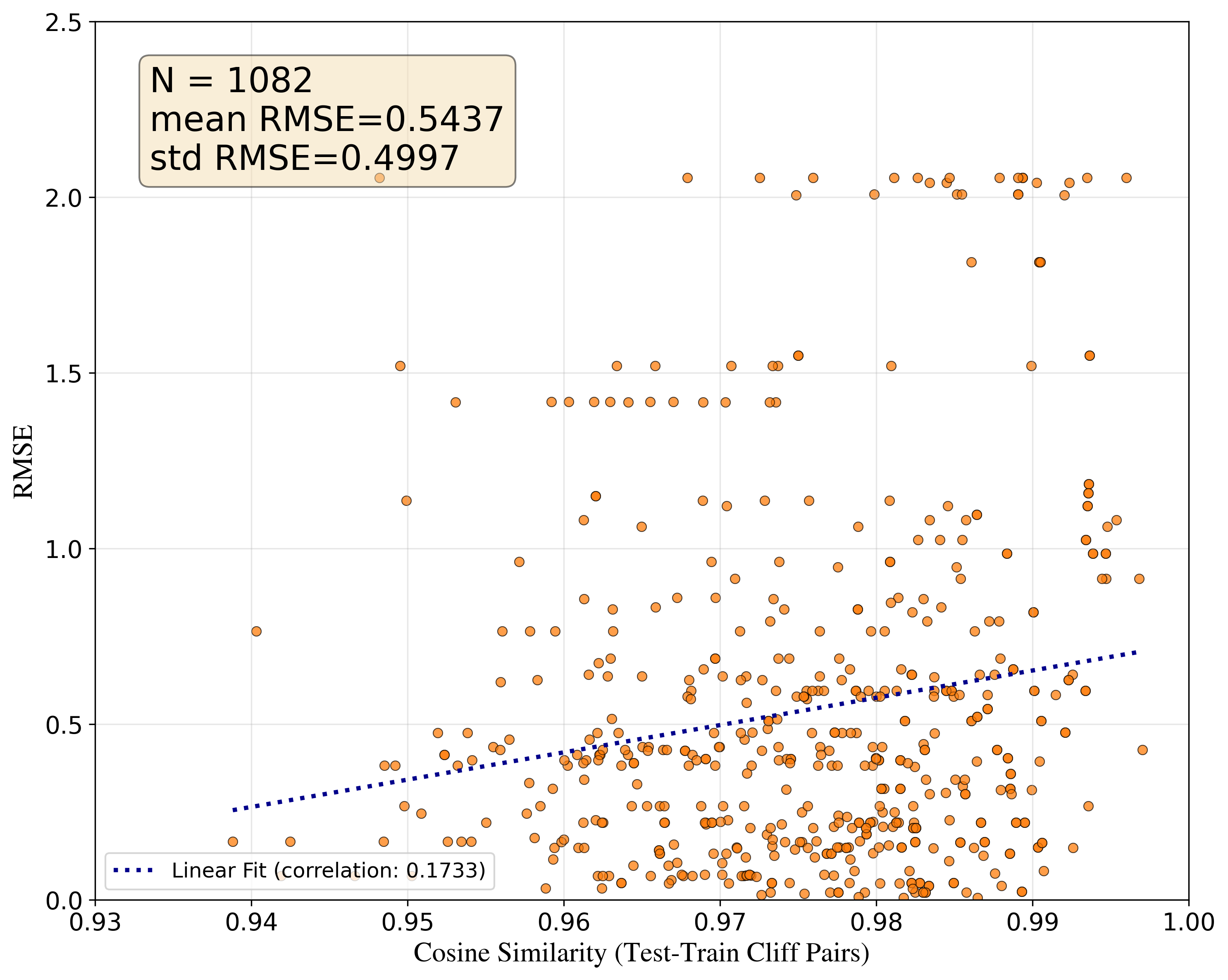}}%
    \subcaptionbox{ECFP}{\includegraphics[width=0.31\linewidth]{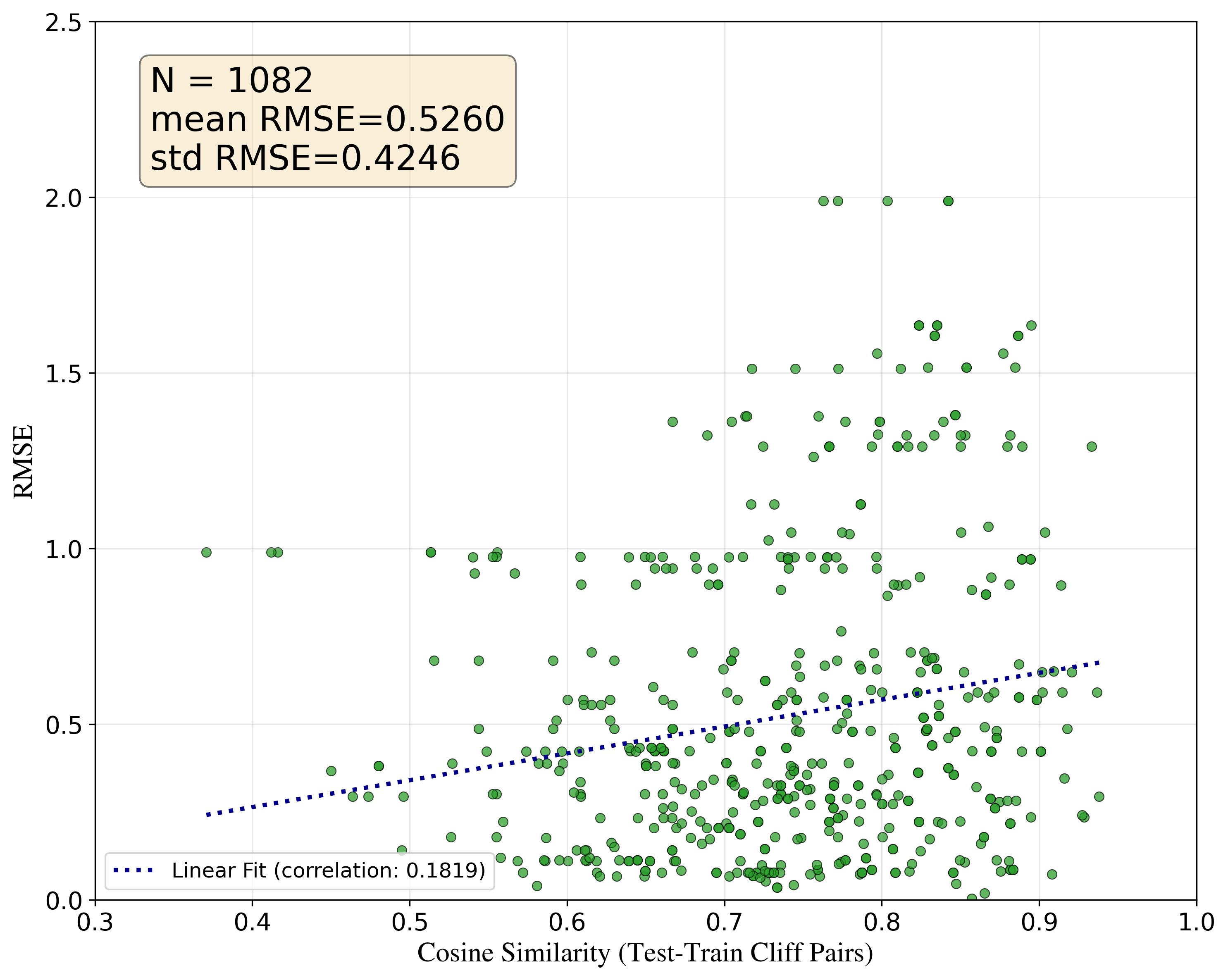}}%
    \subcaptionbox{Chemprop}{\includegraphics[width=0.31\linewidth]{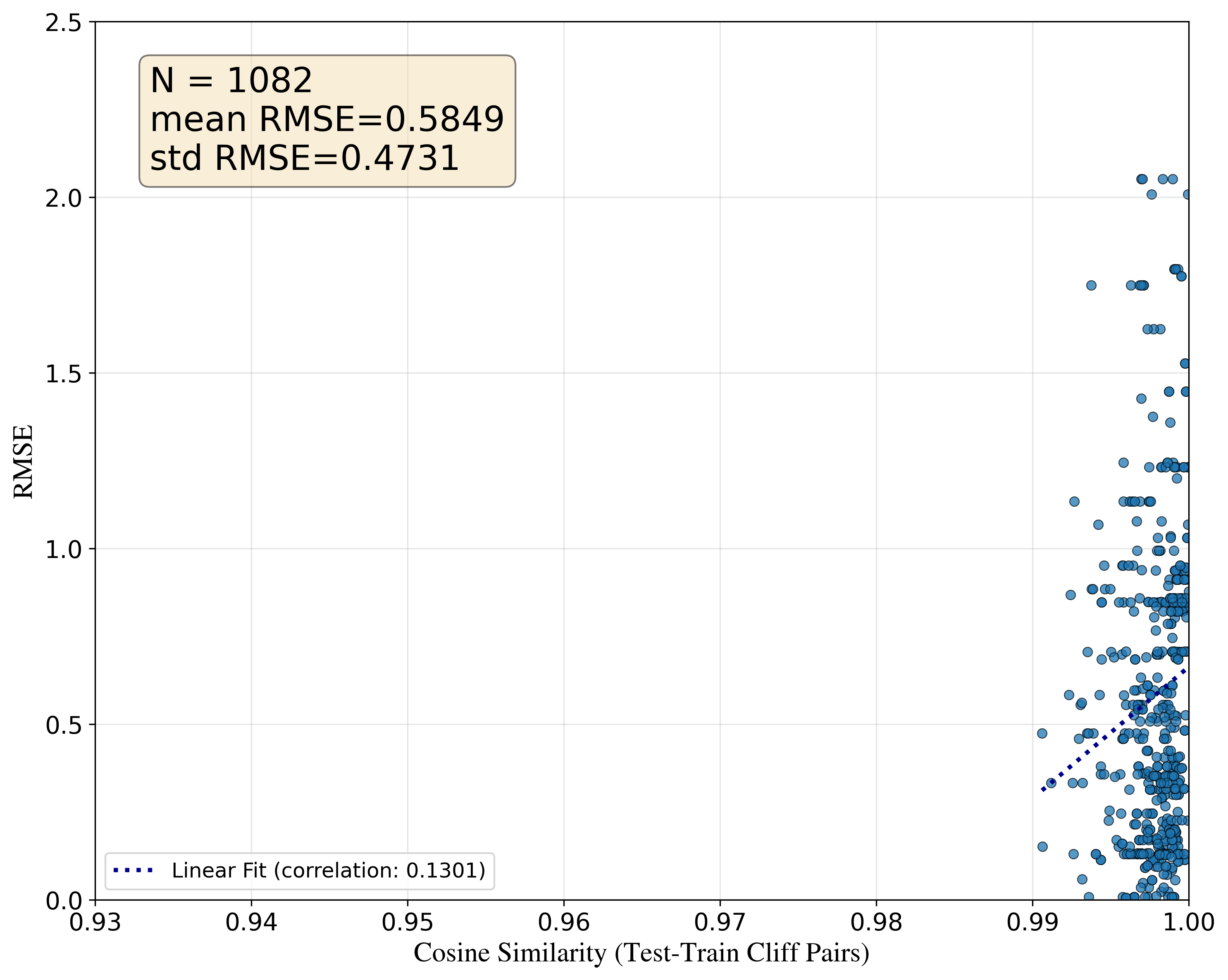}}%
    \captionof{figure}{Correlation between representation similarity and predictive error for activity cliff pairs on CHEMBL4792 (Ki).}%
    \label{fig:error_4792}\par
    \subcaptionbox{GraphCliff}{\includegraphics[width=0.31\linewidth]{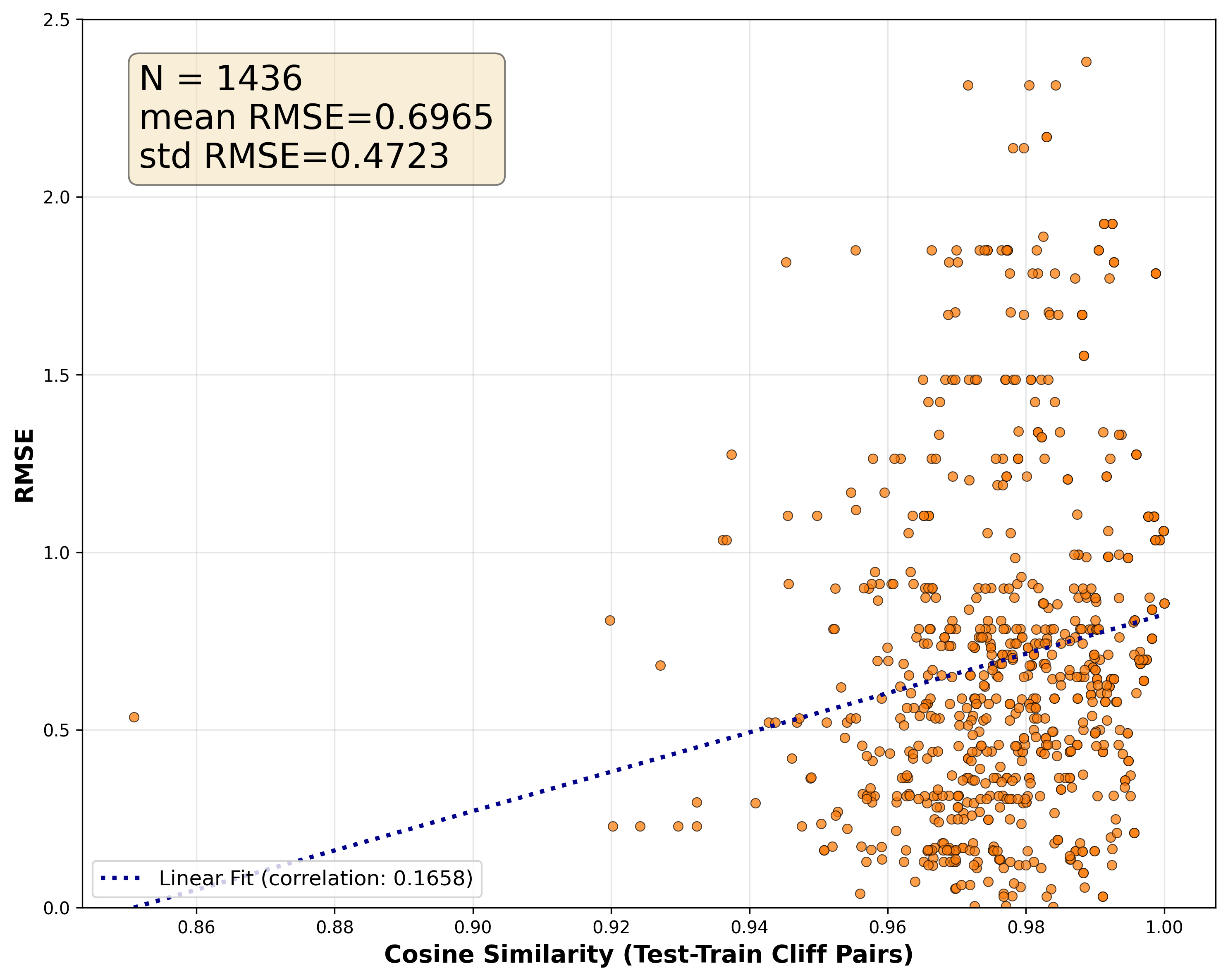}}%
    \subcaptionbox{ECFP}{\includegraphics[width=0.31\linewidth]{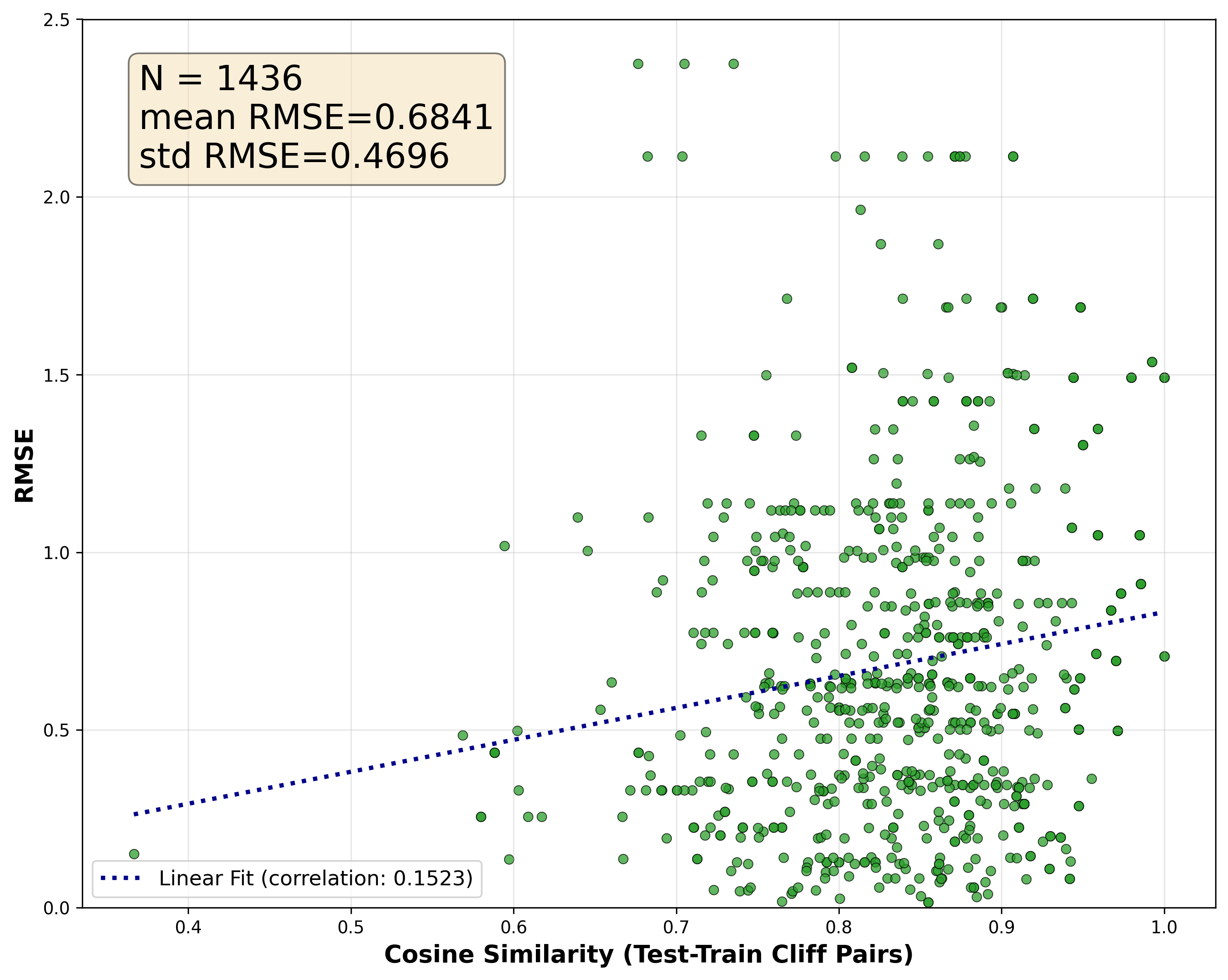}}%
    \subcaptionbox{Chemprop}{\includegraphics[width=0.31\linewidth]{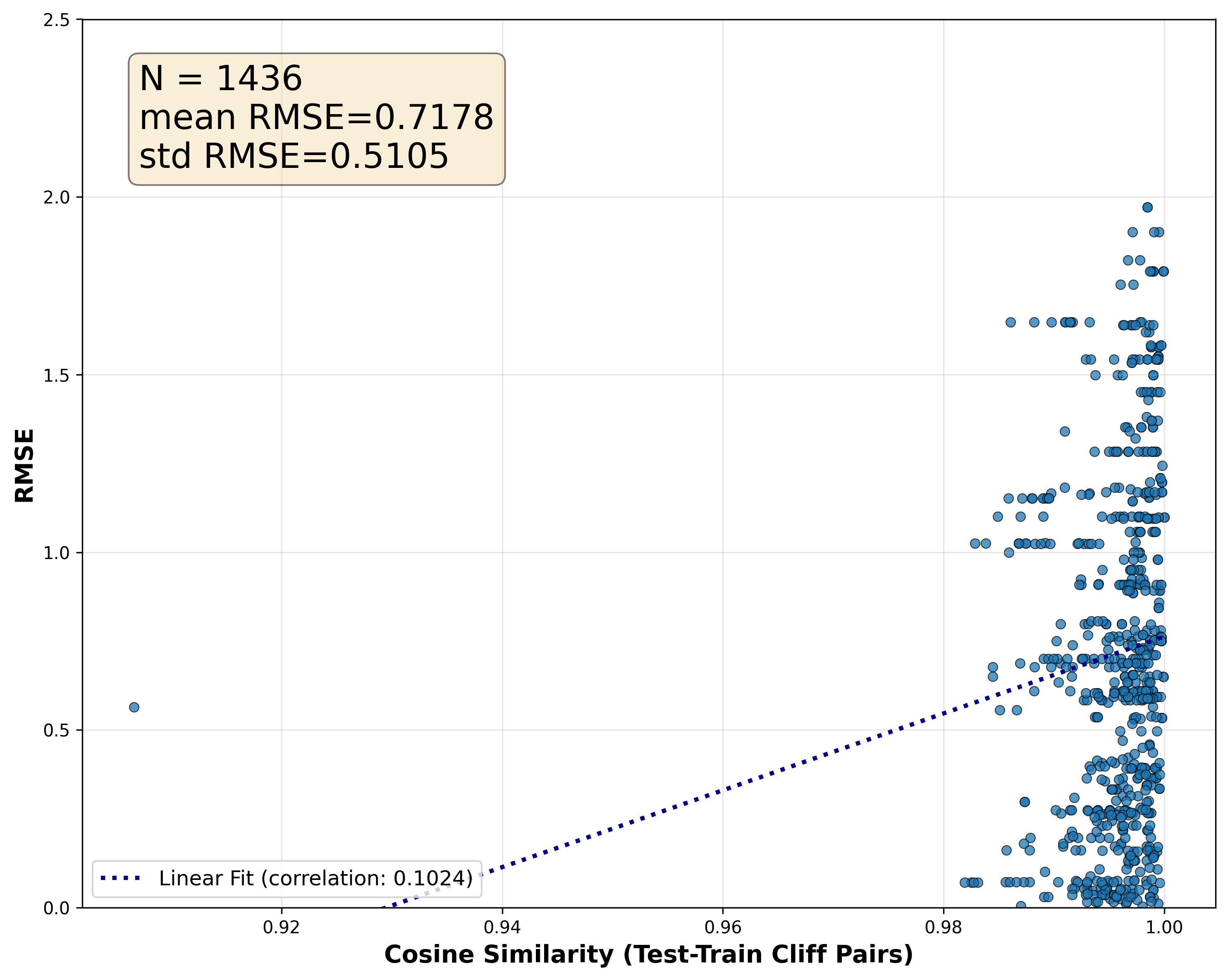}}%
    \captionof{figure}{Correlation between representation similarity and predictive error for activity cliff pairs on CHEMBL237 (Ki).}%
    \label{fig:error_237}
    \end{center}
    \vspace{3em}% gap between the figure block and the two-column body text
]

\subsection{Error Analysis of ECFP-Favored Cases}
\label{sec:error_analysis}

GraphCliff's small residual gap to SVM-ECFP on cliff compounds (mean $\mathrm{RMSE}_{\text{cliff}}$ $0.757$ vs.\ $0.751$) motivates a closer look at where it underperforms. We focus on \emph{train--test} cliff pairs, each a training molecule paired with a structurally similar test molecule of sharply different activity, and relate each pair's cosine similarity under ECFP, GraphCliff, and Chemprop to its prediction error. These similarity scales differ by construction. ECFP is binary and spans $\approx0.3$--$1.0$, whereas the continuous GraphCliff and Chemprop embeddings cluster near $0.9$--$1.0$.

ECFP achieves sharp local discrimination whenever the decisive modification falls within its hashing radius. GraphCliff, by contrast, assigns high similarity to molecules that share broad global structure and so compresses some cliff pairs in embedding space. Even then, its short-range path resolves subtle differences more clearly than Chemprop. The result is a consistent ordering. GraphCliff beats Chemprop but does not always recover the fine-grained distinctions ECFP captures, as seen in the mean error over train--test cliff pairs (measured on cross-split pairs, hence differing from the dataset-level test RMSE), where GraphCliff falls between ECFP and Chemprop on both CHEMBL4792 (Ki) ($0.5437$ vs.\ $0.5260$/$0.5849$; Figure~\ref{fig:error_4792}) and CHEMBL237 (Ki) ($0.6965$ vs.\ $0.6841$/$0.7178$; Figure~\ref{fig:error_237}).

\paragraph{Structural origin of the gap.}
This gap reflects the nature of the two representations rather than insufficient tuning. ECFP hashes each atom-centered substructure to a binary bit, so \emph{any} modification flips a bit and produces maximal local discrimination, a construction-time guarantee that makes ECFP a de facto ``cliff-aware'' descriptor. A learned continuous embedding instead optimizes a regression objective, which exerts a smoothing pressure that draws similar molecules together. GraphCliff narrows this gap relative to every graph baseline but cannot fully close it, because its training signal never explicitly rewards separating cliff pairs.

\paragraph{Directions to narrow the gap.}
Two remedies follow. First, \emph{cliff-aware objectives}: contrastive or margin-based loss terms that maximize embedding separation for cliff pairs would target this similarity compression directly. Second, \emph{hybrid representations}: supplying ECFP as auxiliary features would pair the construction-time discrimination of hashing with GraphCliff's learned, topology-aware embeddings. We regard the former as the more promising near-term direction and leave both to future work.